\documentclass[11pt]{article}

\usepackage[utf8]{inputenc}
\usepackage[T1]{fontenc}
\usepackage{lmodern}
\usepackage[margin=1in]{geometry}
\usepackage{amsmath,amssymb}
\usepackage{graphicx}
\usepackage{booktabs}
\usepackage[round,authoryear]{natbib}
\usepackage{authblk}
\usepackage[hidelinks]{hyperref}

\title{GPROF-IR: An Improved Single-Channel Infrared Precipitation Retrieval for Merged Satellite Precipitation Products}

\author[1]{Simon Pfreundschuh\thanks{Corresponding author: simon.pfreundschuh@colostate.edu}}
\author[1]{Christian D. Kummerow}
\author[2,3]{Jackson Tan}
\author[3]{George J. Huffman}
\affil[1]{Colorado State University, Fort Collins, Colorado, USA}
\affil[2]{University of Maryland, Baltimore County, Baltimore, Maryland, USA}
\affil[3]{NASA Goddard Space Flight Center, Greenbelt, Maryland, USA}

\date{}

\begin{document}

\maketitle

\begin{abstract}
The Integrated Multi-Satellite Retrievals for the Global Precipitation
Measurement (IMERG) product combines precipitation estimates from passive
microwave (PMW) and infrared (IR) sensors to provide quasi-global precipitation
estimates at half-hourly resolution. Because IR observations are primarily
sensitive to cloud-top properties, however, IR-based precipitation estimates are
generally less accurate than PMW estimates and can introduce artifacts into the
merged precipitation fields.

We introduce GPROF-IR, a neural-network-based precipitation retrieval designed
to provide precipitation estimates from single-channel geostationary IR
observations for the upcoming IMERG V08. GPROF-IR uses a convolutional neural
network to exploit spatiotemporal information from sequences of half-hourly IR
observations. We validate GPROF-IR against independent surface-based
precipitation measurements over land and ocean. GPROF-IR substantially improves
upon existing single-channel IR retrievals, including the IR retrieval used in
IMERG V07 and more recent neural-network-based approaches. During overpasses of
flagship PMW sensors, GPROF-IR is more accurate than the corresponding IMERG PMW
estimates over CONUS and Austria, but not Korea. Over CONUS, GPROF-IR
outperforms the merged IMERG V07 product for hourly and daily precipitation
accumulations. Against shipborne disdrometer observations, GPROF-IR is more
accurate than IMERG V07 in the tropics but slightly less accurate in the
extratropics. 

Since geostationary IR observations are continuously available across much of
the globe since 1983, GPROF-IR provides a promising basis for improving merged
precipitation products and developing long-term, temporally consistent
precipitation records.
\end{abstract}

\paragraph{Significance Statement}
Producing global precipitation estimates at high temporal resolution from
satellite observations requires the use of geostationary observations to fill
gaps between overpasses from passive microwave (PMW) sensors in low-earth
orbits. Current global precipitation products use infrared (IR) window-channel
observations that are available at half-hourly resolution from the year 1998
onward. However, since these observations are primarily sensitive to cloud-top
properties, conventional IR-only precipitation algorithms provide only very
crude precipitation estimates, which negatively affects the accuracy of global
precipitation datasets.

We propose a novel machine-learning-based IR retrieval that substantially
improves the accuracy of IR-only precipitation estimates from geostationary
observations. The retrieval is based on a convolutional neural network that
leverages structural as well as temporal information in the IR imagery to
provide more accurate precipitation estimates. Extensive validation against
independent precipitation estimates over both land and ocean demonstrates that
the proposed retrieval can provide more accurate estimates than current merged
precipitation products over CONUS, Austria, and tropical oceans. While the
accuracy of the IR-only retrieval stays below that of merged products over
extra-tropical oceans and Korea, the improvements compared to comparable
retrievals using single-channel IR observations are substantial and robust
across various climate regimes.

The algorithm presented here will be used to produce next version of the
Integrated Multi-Satellite Retrievals for the Global Precipitation Measurement
(IMERG) merged precipitation dataset. It is also fully open source and thus
available to the scientific community to adapt.

%
\section{Introduction}
\label{sec:intro}

The Integrated Multi-Satellite Retrievals for the Global Precipitation
Measurement (IMERG; \citealp{huffman_integrated_2020}) merged precipitation
product is created by combining precipitation estimates derived from two classes
of satellite observations: passive microwave (PMW) measurements from sensors in
low-Earth orbit and infrared (IR) observations from geostationary satellites.
PMW observations provide relatively direct information on surface precipitation
because they measure emission and scattering signals from hydrometeors within
precipitating clouds \citep{kidd_status_2011}. In contrast, IR observations are
primarily sensitive to cloud-top properties and therefore provide only indirect
information about precipitation \citep{tapiador_gpm_2011, kidd_global_2011}.
However, because PMW sensors are carried on satellites in low-Earth orbit, their
observations remain too sparse to provide continuous global coverage at hourly
resolution. Merged precipitation products therefore combine time-propagated PMW
retrievals with precipitation estimates derived from geostationary IR
observations to fill remaining gaps between PMW overpasses.

\begin{figure}[htbp]
 \centerline{\includegraphics[width=37pc]{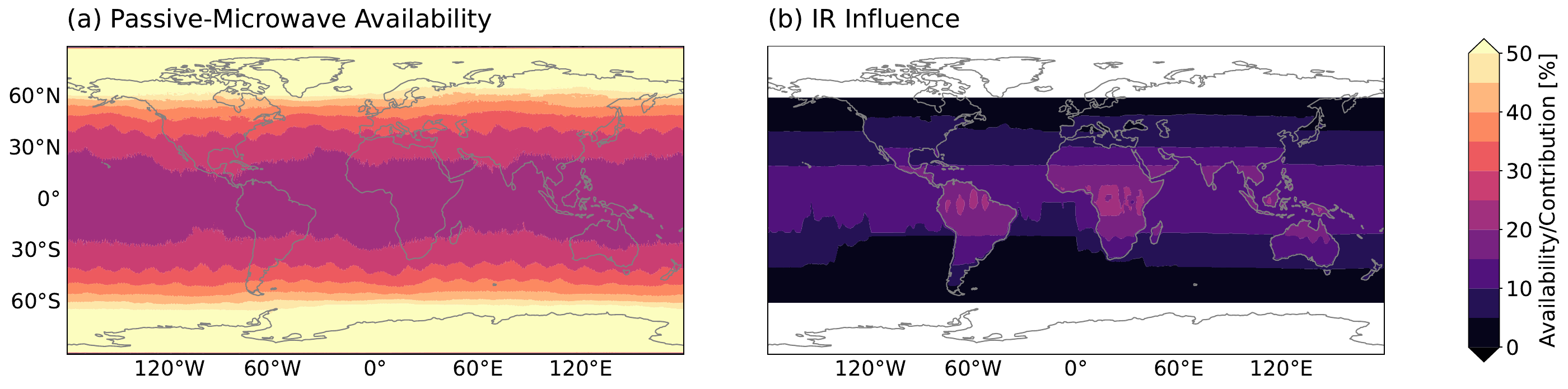}}
 \caption{
Average availability at any one half-hour of passive microwave observations from 2022-01-01 to 2023-01-01 (a) and relative contribution of infrared (IR) observations (b) to IMERG V07 Final Run merged precipitation estimates.
 }\label{fig:availability}
\end{figure}

In practice, the contribution of IR-based precipitation estimates to merged
precipitation products remains limited because current IR retrieval algorithms
exhibit substantially larger uncertainties than PMW-based retrievals. As a
result, error-aware merging frameworks assign only limited weight to IR
precipitation estimates. As illustrated in Fig.~\ref{fig:availability}, for example, the average IR
contribution to 30-minute precipitation estimates in the tropics in IMERG V07
remains below 25\%, despite PMW observations being available only once every two
hours on average. Consequently, the merged precipitation fields rely heavily on
the temporal propagation of time-adjacent PMW retrievals.

\begin{figure}[htbp]
 \centerline{\includegraphics[width=37pc]{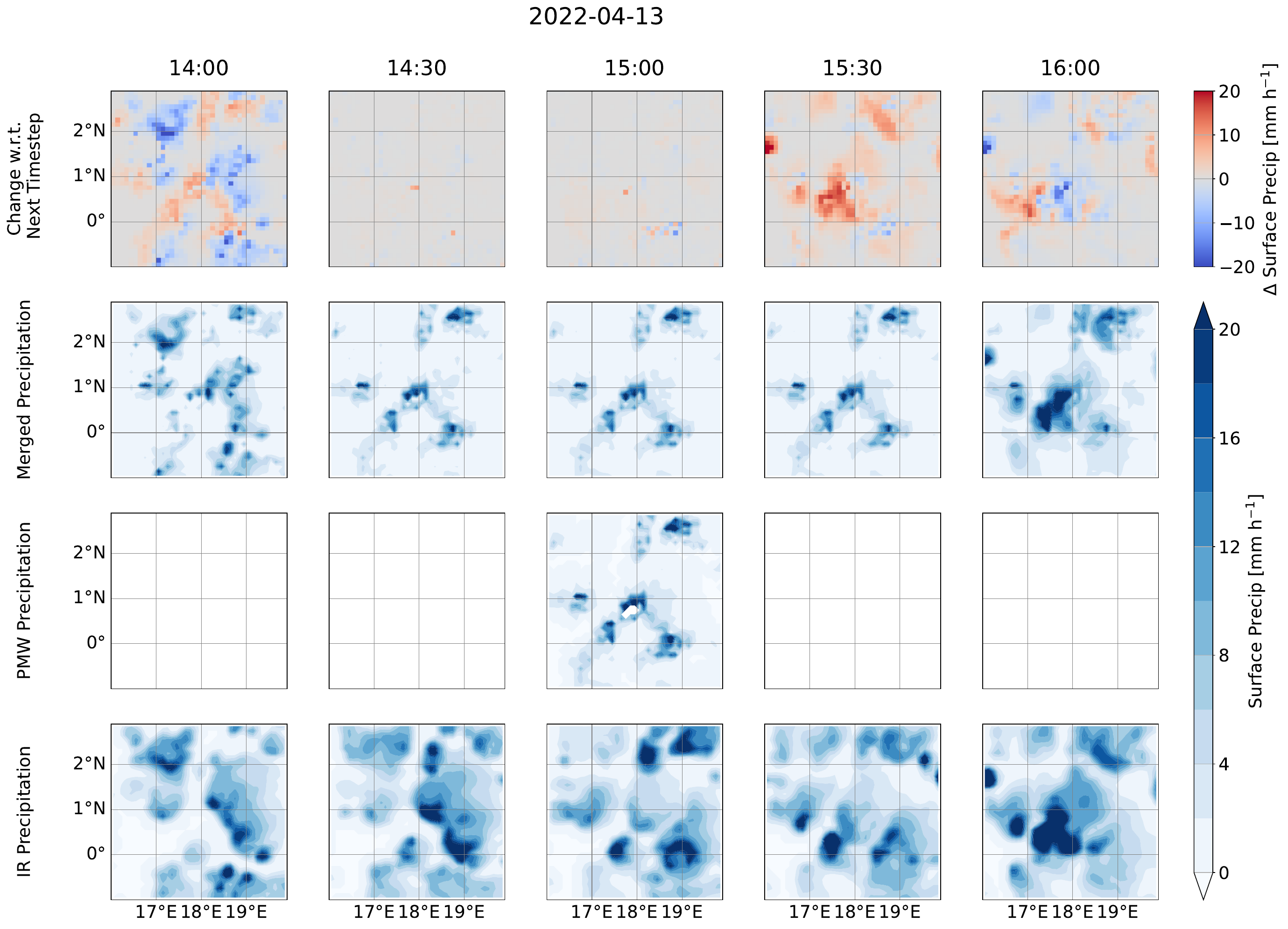}}
 \caption{
    Evolution of convective precipitation cells in IMERG V07. The five
    columns show the half-hourly precipitation field from 14:00 to
    16:00 UTC on April 13, 2022. The first row shows the change in the precipitation
    field relative to the next time step. Rows 2--4 show the final merged IMERG V07
    precipitation field, the passive-microwave-derived precipitation field, and the
    infrared-derived precipitation field, respectively.
 }\label{fig:evolution}
\end{figure}

This dependence on advected PMW estimates can produce physically implausible
representations of rapidly evolving precipitation systems.
Figure~\ref{fig:evolution} highlights this issue for a convective storm case.
While the precipitation field changes markedly between the first and fourth
half-hourly timesteps, it remains nearly static in the intervening period.
Inspection of the corresponding PMW and IR retrievals shows that this stagnation
coincides with a PMW overpass at 15:00 UTC. In IMERG, PMW estimates are given
exclusive weight within a $\pm$30-minute window around such overpasses
\citep{huffman2019imerg_v07_atbd}, such that any variability during this
interval arises solely from advection of the PMW-derived precipitation rates.
Outside this enforced PMW-only window, IR-based estimates begin to contribute to
the merged field. Because the IR and coincident PMW retrievals differ
substantially, their combination introduces abrupt temporal changes. For this
example, the temporal evolution of the merged precipitation is driven more by
inconsistencies between the PMW and IR retrievals than by the underlying
precipitation dynamics. This underscores the need for more accurate IR-based
precipitation estimates to improve the representation of precipitation processes
in merged products.

Due to their continuous coverage over fixed regions of the globe, geostationary
observations have a long history of use for precipitation estimation. Operational
geostationary precipitation algorithms developed over the past several decades
include the GOES Precipitation Index \citep{arkin_87}, Hydro-Estimator
\citep{scofield_2003}, and the self-calibrating multivariate precipitation
retrieval \citep[SCaMPR]{kuligowski_02}. Particularly notable is the PERSIANN
family of retrieval algorithms \citep{hsu_97, hong_04, nguyen_20}, which rely
only on the 11-$\mu$m infrared window channel and can therefore be applied to
the long-term geostationary IR record to produce quasi-global precipitation
estimates extending back to 1983.

Despite early efforts to exploit the spatial structure and temporal evolution of
geostationary observations using hand-engineered features, conventional
single-channel IR retrievals remained fundamentally limited by the ambiguous
relationship between cloud-top infrared radiances and surface precipitation
\citep{nguyen_20}. The widespread adoption of convolutional neural networks
(CNNs) over the past decade has enabled more flexible use of spatial cloud
information directly from satellite observations. Recent studies have shown that
deep-learning-based retrievals that exploit the spatial structure of satellite
imagery can substantially improve precipitation estimates from geostationary
observations \citep{sadeghi2019persiann, amell_probabilistic_2025,
brempong_2025}, even when restricted to a single IR window channel
\citep{pfreundschuh_improved_2022, nguyen_punet_2026}.

Building on these advances, this study introduces the Goddard Profiling
Algorithm Infrared (GPROF-IR), a new retrieval designed to provide improved
IR-based background precipitation estimates for next-generation merged
precipitation products. GPROF-IR uses only the 11-$\mu$m IR window channel,
allowing it to operate within the existing IMERG production framework. In
contrast to existing single-channel IR retrievals, GPROF-IR is trained using
high-quality global reference precipitation estimates and is designed to
maximize the information extracted from geostationary IR observations. Training
targets are derived from designated precipitation sensors to support application
across a broad range of climates and precipitation regimes. The retrieval
ingests IR observations at their native spatial resolution and can incorporate
up to five consecutive 30-minute observations to exploit temporal context. The
complete retrieval and training implementation is openly available, enabling
independent evaluation, reproduction, and further development by the community.

We note that the name GPROF-IR reflects only the intended role of the retrieval
as a complement to GPROF-NN \citep{pfreundschuh_gprof-nn_2022}, which will
provide the PMW precipitation estimates for IMERG V08; it does not imply that
GPROF-IR retrieves vertical hydrometeor profiles.

\section{Methods and Data}
\label{sec:methods}

\subsection{Retrieval Input and Target Data}

\subsubsection{CPC 4-km Gridded IR}

The input data used by the GPROF-IR retrieval are geostationary IR observations
from the NOAA Climate Prediction Center globally merged IR product version 1
\citep{ncep_cpc_gpm_mergir_v1, janowiak2001real}. The CPC IR dataset compiles
geostationary IR observations from the 11-$\mu$m IR window channel from the
global constellation of geostationary satellites on a regular latitude-longitude
grid with a resolution of $0.036^\circ$. The gridded observations at
half-hourly resolution cover much of the globe between $60^\circ$S and
$60^\circ$N from 1998 until the present.

\subsubsection{GPM 2BCMB}

We evaluate two sources of target surface precipitation rates used for training
the GPROF-IR retrieval. The first source is derived from the level-2 combined
radar–radiometer product (2BCMB; \citealp{cmb_data}) of the Global Precipitation
Measurement (GPM; \citealp{hou_global_2014}) mission. The 2BCMB product
integrates observations from the GPM Dual-Frequency Precipitation Radar (DPR)
and the GPM Microwave Imager (GMI) within a variational retrieval framework,
yielding more accurate precipitation estimates than either sensor alone. The
estimates are available continuously from the beginning of the GPM mission in
May 2014 with a spatial sampling of 5 km x 5 km along the DPR swath. The 2BCMB
product serves as the primary reference dataset for training PMW retrievals
across the GPM constellation and is widely understood to provide the most
reliable global precipitation measurements currently available.

\subsubsection{GPROF V08 GMI}

As the second source of reference data, we use GPROF V08 PMW precipitation
retrievals from GMI. GMI is the flagship PMW sensor
of the GPM constellation and combines broad spectral coverage with comparatively
fine spatial resolution. Consequently, GMI-based precipitation retrievals are
among the most accurate PMW precipitation estimates available from the GPM
constellation. In particular, \citet{pfreundschuh_gprof_2024} showed that, when
evaluated against gauge-corrected ground-based radar observations over CONUS,
the accuracy of GMI-based precipitation estimates is comparable to that of the
GPM 2BCMB product. GPROF V08 GMI retrievals are available continuously since the
beginning of the GPM mission, with a spatial sampling of approximately 13~km in
the along-track and 5~km in the across-track direction along the GMI swath.

GPROF V08, the next operational version of the Goddard Profiling Algorithm
(GPROF; \citealp{kummerow2015evolution}), is based on an updated version of the
GPROF-NN 3D retrieval introduced in \citet{pfreundschuh_gprof-nn_2022}. GPROF-NN
3D replaces the Bayesian retrieval scheme used in previous versions of GPROF
with a CNN. Although GPROF V08 is trained primarily using reference
precipitation estimates from the GPM 2BCMB product, several targeted
modifications are applied to mitigate known limitations of these reference data.
These include the use of ground-based radar precipitation estimates over
snow-covered surfaces and the augmentation of the 2BCMB targets over oceans with
precipitation retrieved by the GPROF-NN XPR retrieval
\citep{pfreundschuh_xpr_2026}. GPROF-NN XPR is a GMI-based retrieval
specifically designed to improve the detection and estimation of light and
frozen precipitation that may be missed by the DPR.

\subsection{Baseline Precipitation Estimates}

Because GPROF-IR is intended to provide improved IR-based precipitation
estimates for the upcoming version of IMERG, we compare its retrievals against
both the PMW- and IR-based precipitation estimates contained in the current
IMERG V07 product. Specifically, we use precipitation estimates from the IMERG
Final Run \citep{gpm_imerg_final_v07}. IMERG V07 combines PMW precipitation
estimates from GPROF V07 with IR-based estimates from the PERSIANN Dynamic
Infrared--Rain Rate (PDIR) retrieval \citep{nguyen_20} using a
Kalman-filter-based merging framework \citep{huffman2019imerg_v07_atbd}. Where
possible, we evaluate the PMW- and IR-based estimates separately using the
\textit{MWPrecipitation} and \textit{IRPrecipitation} fields. This allows the
performance of GPROF-IR to be assessed directly against both the current IR
retrieval and the higher-quality PMW estimates that are combined to form the
merged IMERG estimates. IMERG V07 provides quasi-global precipitation estimates
on a 0.1$^\circ$ grid at 30-minute temporal resolution.

We additionally compare against precipitation estimates from the ERA5 reanalysis
\citep{hersbach_era5_2020}. ERA5 provides continuous global precipitation
estimates at hourly temporal resolution on a 0.25$^\circ$ grid, with a record
extending back to 1940. It therefore provides an alternative source of spatially
and temporally continuous precipitation estimates for studying the evolution of
precipitation systems. Unlike satellite retrievals, which infer precipitation
 directly from satellite observations, ERA5 derives precipitation from a
numerical weather prediction model whose atmospheric state is constrained
through the assimilation of a broad range of meteorological observations.
Since ERA5 provides globally complete precipitation estimates continuously from
1940 onward, a novel retrieval intended to improve the representation of
precipitation should provide more accurate estimates than ERA5 to offer clear
added value.

As additional satellite-based baselines, we evaluate PERSIANN-CCS
\citep{hong_04}, PERSIANN-PDIR \citep{nguyen_20}, and PERSIANN-UNet
(P-UNet; \citealp{nguyen_punet_2026}). Like GPROF-IR, these retrievals derive
precipitation primarily from single-channel geostationary IR observations and
therefore provide particularly relevant baselines for assessing the information
extracted from IR imagery. The PERSIANN products are available as hourly
accumulations at a spatial resolution 0.036$^\circ$. To the best of our
knowledge, they are also among the few publicly available long-term
precipitation records based on the 11-$\mu$m
infrared window channel.

Finally, we compare the global precipitation distributions produced by GPROF-IR
with the GPCP V3.3 precipitation climatology \citep{gpcp_v33_precip}. GPCP
combines satellite observations with gauge measurements to provide a long-term,
globally consistent precipitation record designed for climate applications.
Although its comparatively coarse spatial and temporal resolution precludes
direct retrieval-level comparison, its well-established global precipitation
climatology provides an important reference for assessing whether GPROF-IR
reproduces realistic large-scale precipitation distributions. The GPCP V3.3
data is available as daily averages at a spatial resolution of 0.5$^\circ$.

\subsection{Independent Validation Data}

To assess the accuracy and robustness of GPROF-IR, we evaluate the retrieved
precipitation estimates against multiple independent validation datasets.

\subsubsection{The SatRain Benchmark Dataset}

The SatRain benchmark dataset \citep{pfreundschuh2026benchmark} provides a
machine-learning-ready training and evaluation dataset for machine-learning
precipitation retrievals. It integrates high-quality reference data from
gauge-corrected ground radars and the WegenerNet high-density gauge network
\citep{fuchsberger_wegenernet_2021}. The testing split of the benchmark provides
an easily accessible and comprehensive validation resource to for satellite
precipitation retrievals. By using the publicly available SatRain dataset
and following its standardized evaluation protocol, we ensure that the
validation results of the GPROF-IR retrievals are reproducible and directly
comparable to other algorithms. The number of validation samples for the
different testing domains of the SatRain dataset are listed in Table~\ref{tab:satrain_stats}

\begin{table}
  \caption{Sample statistics for the SatRain testing data.}
  \label{tab:satrain_stats}
  \centering
  \begin{tabular}{lrrr}
    \hline
    Domain &
    Total samples &
    $P > 0.1$ mm h$^{-1}$ &
    $P > 10$ mm h$^{-1}$ \\
    \hline
    Austria & 19\,121      & 1916 & 4 \\
    CONUS   & 754\,130\,707 & 51\,097\,652 & 1\,001\,067 \\
    Korea   & 27\,355\,696  & 2\,946\,193 & 92\,496 \\
    \hline
  \end{tabular}
\end{table}

The analysis presented here focuses on the GMI-based testing data, which
comprises one year of GMI overpasses over CONUS and Korea as well as two years
of overpasses over the Wegener gauge network in Austria. The CONUS testing data
is derived from gauge-corrected precipitation estimates from the NOAA
Multi-Radar Multi-Sensor (MRMS, \citealp{smith_multi-radar_2016}) products; the
Korea data is derived from gauge-corrected precipitation estimates provided by
the Korean Meteorological Agency \citep{ryu_radar_2025}. To accommodate for the
testing resolution of 0.036 degree used by the SatRain dataset, all GPROF-IR and
baseline estimates are interpolated to the SatRain grid using bilinear
interpolation.

The SatRain evaluation protocol considers only pixels for which valid,
collocated PMW observations are available. Consequently, all IMERG PMW estimates
are derived from GMI observations for the GMI-based SatRain test data used here.
Because GMI is among the sensors with the highest resolution and spectral
coverage in the GPM constellation, this comparison represents a particularly
strong PMW baseline.

\subsubsection{OceanRAIN Shipborne Distrometer Measurements}

\begin{figure}[htbp]
 \centerline{\includegraphics[width=37pc]{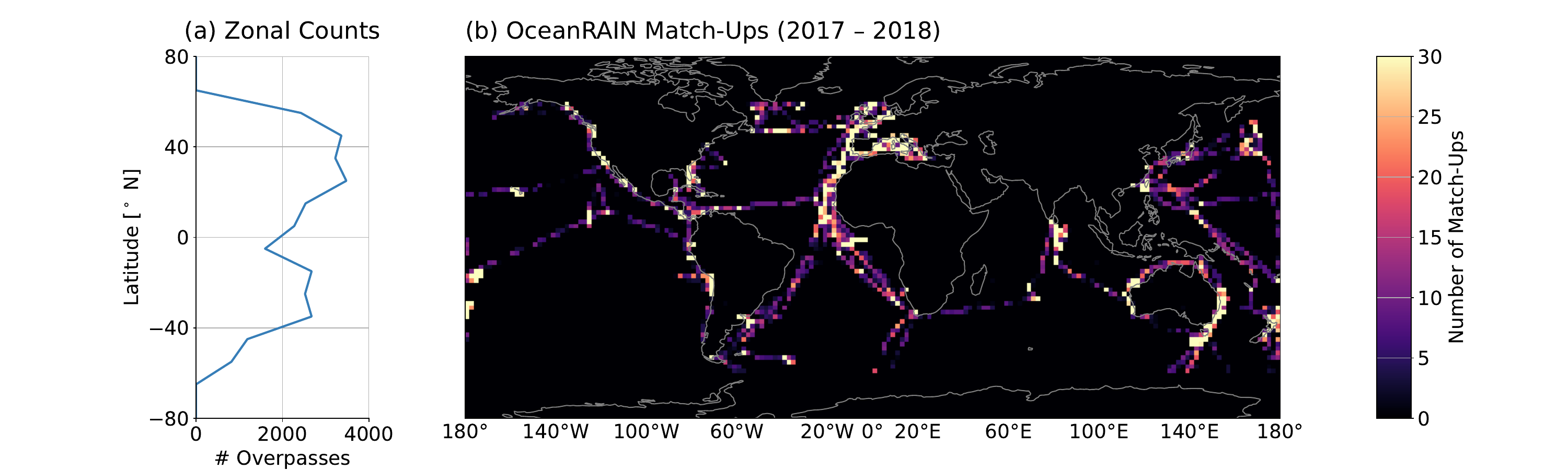}}
 \caption{
   Zonal (a) and global (b) counts of matched OceanRAIN shipborne distrometer
   estimates and corresponding satellite precipitation estimates used to
   validate the GPROF-IR retrieval.
 }\label{fig:ocean_rain}
\end{figure}

Since the SatRain dataset relies on measurements from gauges and gauge-corrected
ground-based radars, it is mostly limited to land surfaces. To also validate the
retrievals over ocean, we use shipborne distrometer measurements from the
OceanRAIN dataset \citep{klepp2018oceanrain} from all measurements available
during the years 2017 and 2018. The OceanRAIN distrometer measurements are
provided at 1-min temporal resolution from which we derive hourly accumulations.
The corresponding hourly accumulations from satellite and reanalysis data are
matched to the shipborne accumulations by interpolating the former bilinearly along the
ship-track. The spatial distribution of the matchups between OceanRAIN, GPROF-IR
and all baseline datasets are shown in Fig.~\ref{fig:ocean_rain}. Since the
validation period overlaps with the training data, we exclude all measurements
within six hours of a GMI overpass over any of the OceanRAIN vessels. Since the
GPROF-IR training targets are both derived from observations from the GPM
satellite, this excludes all precipitation scenes that the retrieval model has
encountered during training with ample margin and should avoid data leakage and
thus potential biases in the validation. After excluding GMI overpasses and
discarding invalid measurements in the distrometer measurements or any of the
retrievals, the OceanRAIN validation dataset comprises 28768 measurement hours
across seven different vessels. We further stratify the validation data into
measurements in the tropics ($|\theta| < 30^\circ$) and extratropics
($30 \leq \theta$) with 15076 and 13693 measurement hours, respectively.

\subsubsection{Hourly and Daily Precipitation Accumulations over CONUS}

Because the SatRain test data are restricted to individual PMW overpasses over
the evaluation domains, their temporal sampling is too sparse to reliably assess
precipitation accumulations or spatial variations in retrieval accuracy. We
therefore complement the SatRain evaluation over CONUS with an analysis of
hourly and daily precipitation accumulations. This evaluation also provides a
more appropriate basis for assessing the PERSIANN baseline retrievals, which are
available only as hourly accumulations. As reference data, we use hourly
gauge-corrected precipitation estimates from MRMS from the full year 2022,
aggregated to a 0.1$^\circ$ grid to match the spatial sampling of IMERG. To
restrict the evaluation to locations with reliable radar coverage, we use the
MRMS Radar Quality Index (RQI) from 30 minutes after each hour and retain only
pixels with RQI values exceeding 0.5. Hourly accumulations from the
higher-resolution satellite retrieval products are matched to the MRMS estimates
by spatially averaging them onto the 0.1$^\circ$ grid. IMERG V07 and ERA5
estimates are mapped to the 0.1$^\circ$ grid using bilinear interpolation.

\subsection{Implementation}
\subsubsection{Neural Network Model}

\begin{figure}[htbp]
 \centerline{\includegraphics[width=37pc]{figures/figure04.pdf}}
 \caption{
Neural network architecture of the GPROF-IR retrieval. The U-Net is constructed from fused MBConv blocks (MBConvF) and MBConv blocks, with $e$ denoting the expansion factor used in the inverted bottleneck, following the EfficientNet-V2 design of \citet{tan2021efficientnetv2}.
 }\label{fig:architecture}

\end{figure}

Note that \textit{model} is used in this paper to denote a neural network model,
not to be confused with a numerical model of precipitation, weather, or climate
processes.

The GPROF-IR retrieval is based on the same U-Net \citep{ronneberger_u-net_2015}
architecture used for the GPROF V08 retrieval. The network – as illustrated in
Fig.~\ref{fig:architecture} – is composed of EfficientNet-V2–style inverted bottleneck convolutional
blocks \citep{tan_efficientnetv2_2021}. Following the normalization and
activation design adopted in ConvNeXt \citep{liu_convnet_2022}, each
convolutional block uses Layer Normalization \citep{ba2016layer} and GELU
activation functions \citep{hendrycks2016gaussian}. The network architecture was
not subject to an extensive architecture search; instead, its design combines
established components from modern convolutional neural networks with experience
gained from the development of previous ML-based precipitation retrievals
\citep{pfreundschuh_improved_2022, pfreundschuh_gprof-nn_2022,
amell_chalmers_2023, amell_probabilistic_2025}. Compared to the GPROF V08
models, the network was reduced in both width and depth to meet operational
processing-time requirements. The architecture of the resulting model is
summarized in Table~\ref{tab:architecture}.

\begin{table}
  \caption{
    Network architecture of the convolutional neural network used for GPROF-IR.
    Here, $k$ denotes the convolutional kernel size and $e$ the expansion factor
    of the inverted bottleneck in the MBConv blocks.
  }
  \label{tab:architecture}
  \centering
  \small
  \setlength{\tabcolsep}{5pt}
  \begin{tabular}{lclrrc}
    \hline
    Component &
    Depth &
    Block &
    \shortstack{Input\\channels} &
    \shortstack{Output\\channels} &
    \shortstack{Spatial\\size} \\
    \hline
    Stem              & 1 & Simple Convolution, $k=3$                         & 1, 3, or 5 & 29  & $512 \times 512$ \\
    Encoder, stage 1  & 1 & Fused MBConv, $k=3$, $e=1$         & 29         & 29  & $512 \times 512$ \\
    Encoder, stage 2  & 1 & Fused MBConv, $k=3$, $e=4$         & 29         & 58  & $256 \times 256$ \\
    Encoder, stage 3  & 2 & Fused MBConv, $k=3$, $e=4$         & 58         & 77  & $128 \times 128$ \\
    Encoder, stage 4  & 2 & MBConv, $k=3$, $e=4$               & 77         & 154 & $64 \times 64$ \\
    Encoder, stage 5  & 3 & MBConv, $k=3$, $e=6$               & 154        & 192 & $32 \times 32$ \\
    Encoder, stage 6  & 4 & MBConv, $k=3$, $e=6$               & 192        & 307 & $16 \times 16$ \\
    \hline
    Decoder, stage 1  & 3 & MBConv, $k=3$, $e=6$               & $307+192$  & 192 & $32 \times 32$ \\
    Decoder, stage 2  & 2 & MBConv, $k=3$, $e=4$               & $192+154$  & 154 & $64 \times 64$ \\
    Decoder, stage 3  & 1 & Fused MBConv, $k=3$, $e=4$         & $154+77$   & 77  & $128 \times 128$ \\
    Decoder, stage 4  & 1 & Fused MBConv, $k=3$, $e=4$               & $77+58$    & 58  & $256 \times 256$ \\
    \hline
    Head              & 3 & Conv, Layer Normalization, GELU, $k=1$       & 58         & 32  & $256 \times 256$ \\
    \hline
  \end{tabular}
\end{table}

All GPROF-IR models were trained for 140 epochs using the AdamW optimizer
\citep{loshchilov_decoupled_2019} with an initial learning rate of
$5 \cdot 10^{-4}$. A cosine-annealing learning-rate schedule with warm restarts
\citep{loshchilov2016sgdr} after 20 and 60 epochs was used to improve
convergence during training.

\subsubsection{Retrieval Input and Output}

The retrieval input consists of single-channel infrared observations from up to
five consecutive timesteps leading up to the nominal time. Preceding timesteps
were used instead of centered timesteps to avoid impacting the latency of
near-real-time retrievals. To leverage the comparably high resolution of the IR
input observations, the model ingests the gridded IR observations at their
native 0.036$^\circ$ and 30 min resolution. Since the effective spatial
resolution of the retrieved precipitation can be expected to be much lower than
the 0.036$^\circ$ of the input observations, the model output is produced at
half of the input spatial resolution to reduce inference complexity as well as
storage requirements. The input consists of image patches of size 512 px x 512
px at 0.036$^\circ$-resolution that are mapped to precipitation estimates of
size 256 px $\times$ 256 px with a resolution of 0.072$^\circ$. The model
outputs are therefore at a spatial resolution of 0.072$^\circ$ and a temporal
resolution of 30 minutes.

\subsubsection{Training Data}

The training dataset consists of IR observations paired with corresponding
reference precipitation estimates within $\pm$15 minutes of the nominal
observation time. The model is trained using all available reference data from
2015–2018, with data from January and June 2020 used for validation during
training and temporally independent data from 2022 are used to determine the
test-set accuracy reported below. The network is trained to predict 32 quantiles
of the posterior distribution of the target precipitation rate using a
quantile-regression loss \citep{pfreundschuh_neural_2018}. The principal
retrieval output is the mean of the posterior distribution obtained from these
32 quantiles by approximating the cumulative distribution function and
calculating its mean. The resulting posterior mean is approximately bias free
and minimizes the mean-squared error of retrieval output. However, the
probabilistic nature of the network also allows the calculation of other
statistical properties of the posterior distribution such as exceedance
probability for arbitrary precipitation rates, which we will use later to assess
the precipitation detection capabilities of the model.

\subsubsection{Design Experiments}

Gridded geostationary IR observations provide only indirect information on
surface precipitation, as they are primarily sensitive to cloud-top properties.
However, their high temporal resolution captures the evolution of cloud systems,
which may help constrain precipitation at the surface. To exploit this temporal
information, we evaluate retrieval configurations that incorporate multiple
preceding IR timesteps. Specifically, we consider models that include two and
four additional prior observations (for a total of three and five timesteps,
respectively). These multi-timestep models are trained sequentially: the
single-timestep model is trained first, its weights are used to initialize the
three-timestep model, and the latter is subsequently used to initialize the
five-timestep model.

Overall, we test the following retrieval configurations:

\begin{description}
    \item[GPROF-IR (CMB, 1)] uses input observations from a single time step and
      is trained using the 2BCMB dataset as reference data.
    \item[GPROF-IR (GMI, 1)] uses input observations from a single time step and
      trained using GPROF V08 GMI retrieval results as reference data.
    \item[GPROF-IR (GMI, 3)] uses input observations from three time steps and
      trained using GPROF V08 GMI retrieval results as reference data.
    \item[GPROF-IR (GMI, 5)] uses input observations from five time steps and
      trained using GPROF V08 GMI retrieval results as reference data.
\end{description}

\subsection{Metrics}

The following metrics will be used to assess the GPROF-IR retrievals.

\subsubsection{Precipitation Quantification}

The bias is the systematic error between the estimated precipitation $\hat{y}$
and the reference value $y$. A bias of $0 \%$ indicates that the
retrieval produces the same amount of precipitation as the reference estimates
across all evaluated test scenes. The bias is defined here so that a positive
bias signifies overestimation of precipitation and a negative bias
underestimation. We report the mean in percent relative to the mean of the
reference data. The bias is calculated using

\begin{align}
  \mathrm{Bias} &= 100 \cdot \sum_{i=1}^{N}\frac{(\hat{y}_i-y_i)}{y_i},
\end{align}
where the sum is calculated over the validation samples $y_1, \ldots, y_N$.

The mean absolute error (MAE) is the expected absolute error of the retrieval. An error-free retrieval has an absolute error of 0.

\begin{align}
\mathrm{MAE} &= \frac{1}{N}\sum_{i=1}^{N}\left|\hat{y}_i-y_i\right|
\end{align}

The mean squared error (MSE) is the expected squared error between prediction and reference. An  error-free retrieval has a mean-squared error of 0.

\begin{align}
\mathrm{MSE} &= \frac{1}{N}\sum_{i=1}^{N}(\hat{y}_i-y_i)^2
\end{align}

The correlation coefficient $r$ quantifies the co-variability between the precipitation estimates and reference data. An error-free retrieval has a correlation coefficient of 1 while random guesses would result in a correlation coefficient of 0:

\begin{align}
r &= \frac{\sum_{i=1}^{N}(\hat{y}_i-\bar{\hat{y}})(y_i-\bar{y})}{\sqrt{\sum_{i=1}^{N}(\hat{y}_i-\bar{\hat{y}})^2 \sum_{i=1}^{N}(y_i-\bar{y})^2}}
\end{align}

While the point-wise metrics described above are sensitive to errors in the
spatial distribution of precipitation, they do not explicitly characterize the
spatial scales represented by the retrieval. We therefore complement the
point-wise evaluation with an analysis of the effective resolution, which
quantifies the finest spatial scale at which the retrieval retains meaningful
agreement with the reference precipitation fields. Here, we follow the formulation
introduced in \citet{pfreundschuh_gprof_2024}, which is based on the work by
\citet{guilloteau_multiscale_2020} and \citet{guilloteau2021well}. The effective
resolution is derived from the spectral coherence $C_{\hat{y},y}$ between the retrieved precipitation field $\hat{y}$ and and reference precipitation field $y$:

\begin{align}
C_{\hat{y},y}(k) &= \frac{\left|P_{\hat{y},y}(k)\right|} {\sqrt{P_{\hat{y}}(k)P_y(k)}},
\end{align}

where \(P_{\hat{y}}\), \(P_y\), and \(P_{\hat{y},y}\) are the radially averaged auto- and
cross-power spectral densities
\begin{align}
P_{\hat{y}}(k) &= \left\langle  \text{DCT}(\hat{y})(k_x,k_y) \cdot \text{DCT}(\hat{y})(k_x,k_y)^{*}\right\rangle, \\
P_{y}(k) &= \left\langle \text{DCT}(y)(k_x,k_y) \cdot \text{DCT}(y)(k_x,k_y)^{*}\right\rangle, \\
P_{\hat{y},y}(k) &= \left\langle  \text{DCT}(\hat{y})(k_x,k_y) \cdot \text{DCT}(y)h(k_x,k_y)^{*}\right\rangle.
\end{align}

Here we use $\text{DCT}(\cdot)$ to denote the coefficients of a 2D discrete cosine transform
of a given retrieved or reference precipitation field and $^{*}$ to denote the
complex conjugate. The averages $\langle\cdot\rangle$ are computed over
non-overlapping windows of size $32\ \text{px} \times 32\ \text{px}$ containing
no invalid retrieved or reference precipitation estimates and over all 2D DCT coefficients
with the same radial wavenumber.

We then define the effective resolution as half the wavelength of the finest wavenumber $k$ at which the spectral coherence exceeds $\frac{1}{\sqrt{2}}$:

\begin{align}
\frac{1}{\sqrt{2}} \leq C_{\hat{y},y}(k)
\end{align}

The effective resolution thus corresponds to the finest spatial
scale at which the coefficient of determination between retrieved and
actual DCT coefficients exceeds 0.5.

\subsubsection{Precipitation Detection}

To complement the assessment of the quantitative precipitation estimation skill, we also asses the ability
to detect  precipitation and heavy precipitation using the following detection metrics. Given a binary
true-or-false prediction $\hat{y}$ of the manifested event $y$, we define the following quantities:

\begin{description}
      \item[True Positives (TP)]: the total number of samples for which both prediction $\hat{y}$ and manifested event $y$ are true
      \item[False Positives (FP)]:  the total number of samples for which the prediction $\hat{y}$ is true but the event $y$ is false
  \item[True Negatives (TN)]:  the total number of samples for which both prediction $\hat{y}$ is false and the manifested event is false
  \item[False Negatives (TN)]:  the total number of samples for which the  prediction $\hat{y}$ was false but the event $y$ is true
\end{description}

We use the Heidke-Skill Score (HSS) to quantify the detection skill of the
retrievals. The HSS quantifies the skill of the classifier relative to a random
reshuffling of the predictions. A perfect detector corresponds to
$\text{HSS} = 1$ , whereas $\text{HSS} = 0$ corresponds to random guessing using
the occurrence frequencies of the predictor.

\begin{align}
\text{HSS} &= \frac{2\left(\mathrm{TP}\cdot\mathrm{TN} - \mathrm{FP}\cdot\mathrm{FN} \right)}{(\mathrm{TP}+\mathrm{FN})\cdot(\mathrm{FN}+\mathrm{TN}) + (\mathrm{TP}+\mathrm{FP})\cdot(\mathrm{FP}+\mathrm{TN})}
\end{align}

\section{Results}
\label{sec:results}

Below we asses the retrieval accuracy of GPROF-IR, validate it using independent
precipitation estimates, and characterize the global precipitation climatology
from a year of retrievals.

\subsection{Test-Set Accuracy}

\begin{figure}[htbp]
 \centerline{\includegraphics[width=37pc]{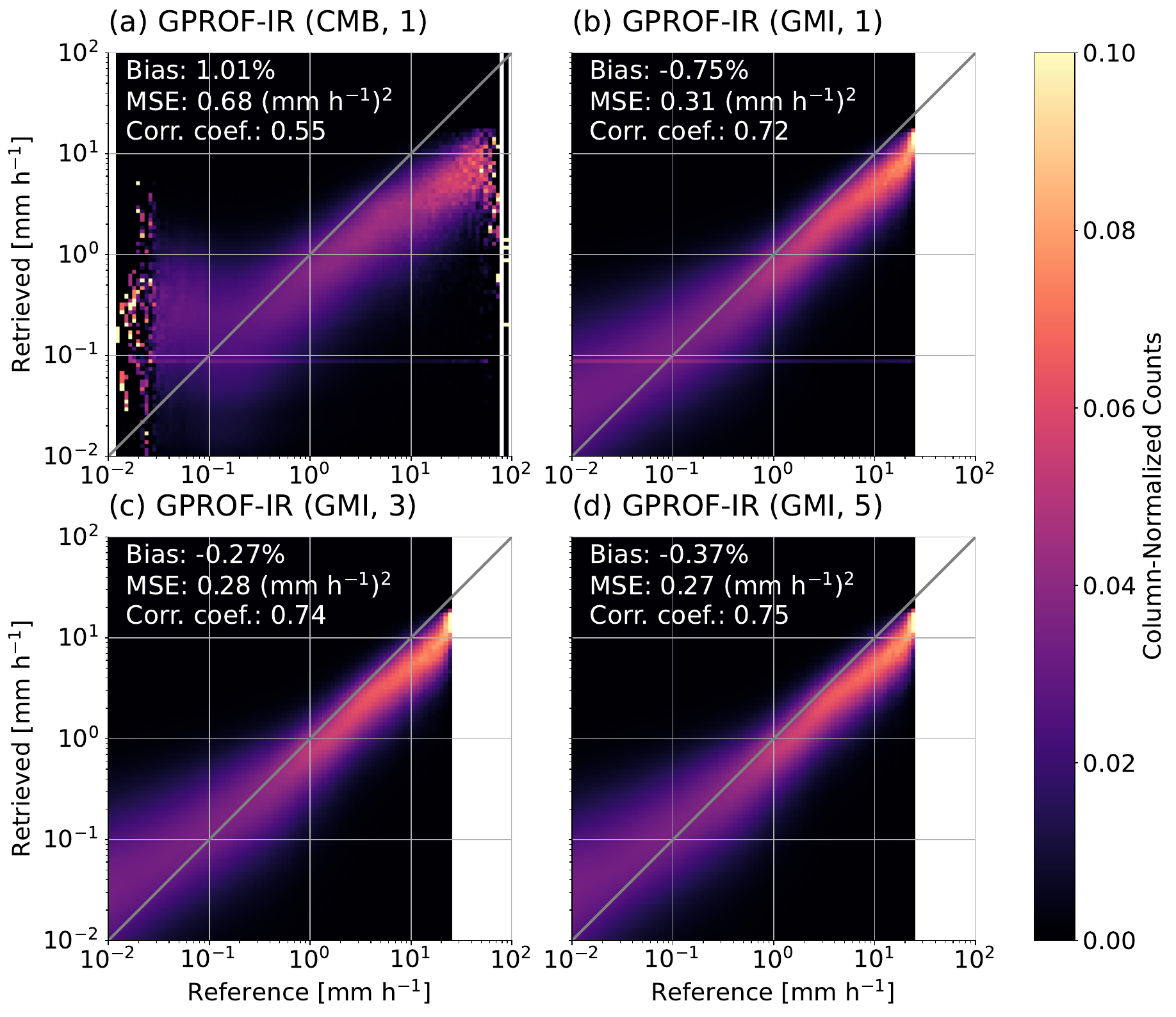}}
 \caption{
Test-set accuracy of the GPROF-IR retrieval. Each panel shows a scatter plot
showing the relation between reference and retrieved precipitation for (a) the
single-step retrieval using GPM 2BCMB reference data, (b) the single-step
retrieval using GPROF-NN GMI reference data, (c) a three-step retrieval using
GMI reference data, and (d) a 5-step retrieval using GMI reference data.
 }\label{fig:test_results}
\end{figure}

We begin by assessing the retrieval accuracy on temporally independent testing
data from the year 2022. Similar to the training data, the testing data for the
different GPROF-IR configurations uses different target precipitation estimates.
The testing data for the GPROF-IR (CMB, 1) configuration uses reference estimates
from 2BCMB whereas the other configurations use estimates from GPROF V08 GMI.

Figure~\ref{fig:test_results} displays scatter plots and accuracy statistics for
the four tested retrieval configurations. All four retrievals demonstrate skill
in reproducing the reference precipitation estimates with low biases and
mean-squared error and high correlation coefficients. However, they also exhibit
a tendency to overestimate light precipitation and underestimate heavy
precipitation, which is typically observed for precipitation retrievals from
passive sensors. The GPROF-IR (CMB, 1) retrieval exhibits notably larger spread
than the GPROF V08 GMI-based retrievals, which is a result of the higher spatial
resolution of the 2BCMB reference data. The three-timestep retrieval GPROF-IR
(GMI, 3) shows small improvements over the single-timestep GMI retrieval
indicating that it is able to leverage temporal information in the IR imagery to
better constrain precipitation estimates. The GPROF-IR (GMI, 5) retrieval shows
further, but even smaller improvements over the three-step retrieval, indicating
that the information extracted from the temporal evolution may begin to
saturate.

The spatial distributions of bias, MAE, MSE, and correlation are shown in
Fig.~\ref{fig:test_results_spatial} in the appendix. All retrievals exhibit notable wet biases over
climatologically dry regions, particularly in the subtropics. This likely
reflects difficulties in distinguishing precipitating from non-precipitating
clouds, an effect that is amplified by normalization with the reference
climatology. The wet bias is strongest for the GPROF-IR (CMB, 1) retrieval and
is substantially reduced for the GMI-based retrievals. The spatial distributions
of MAE and MSE largely follow the global distribution of precipitation, with the
GMI-based retrievals exhibiting generally lower errors, consistent with the
lower effective resolution of their reference data. In contrast, the correlation
coefficient exhibits comparatively little spatial variability and remains high
across most of the globe, with the principal exception of the subtropical
subsidence regions. This indicates that the retrievals reproduce much of the
spatial and temporal variability in precipitation across most regions but
perform less well in the subtropics, where shallow and warm-rain processes are
more prevalent. Such precipitation is inherently difficult to diagnose from IR
observations alone, although the reduced agreement may also partly reflect
greater uncertainty in the reference data in these regions.

\begin{figure}[htbp]
 \centerline{\includegraphics[width=37pc]{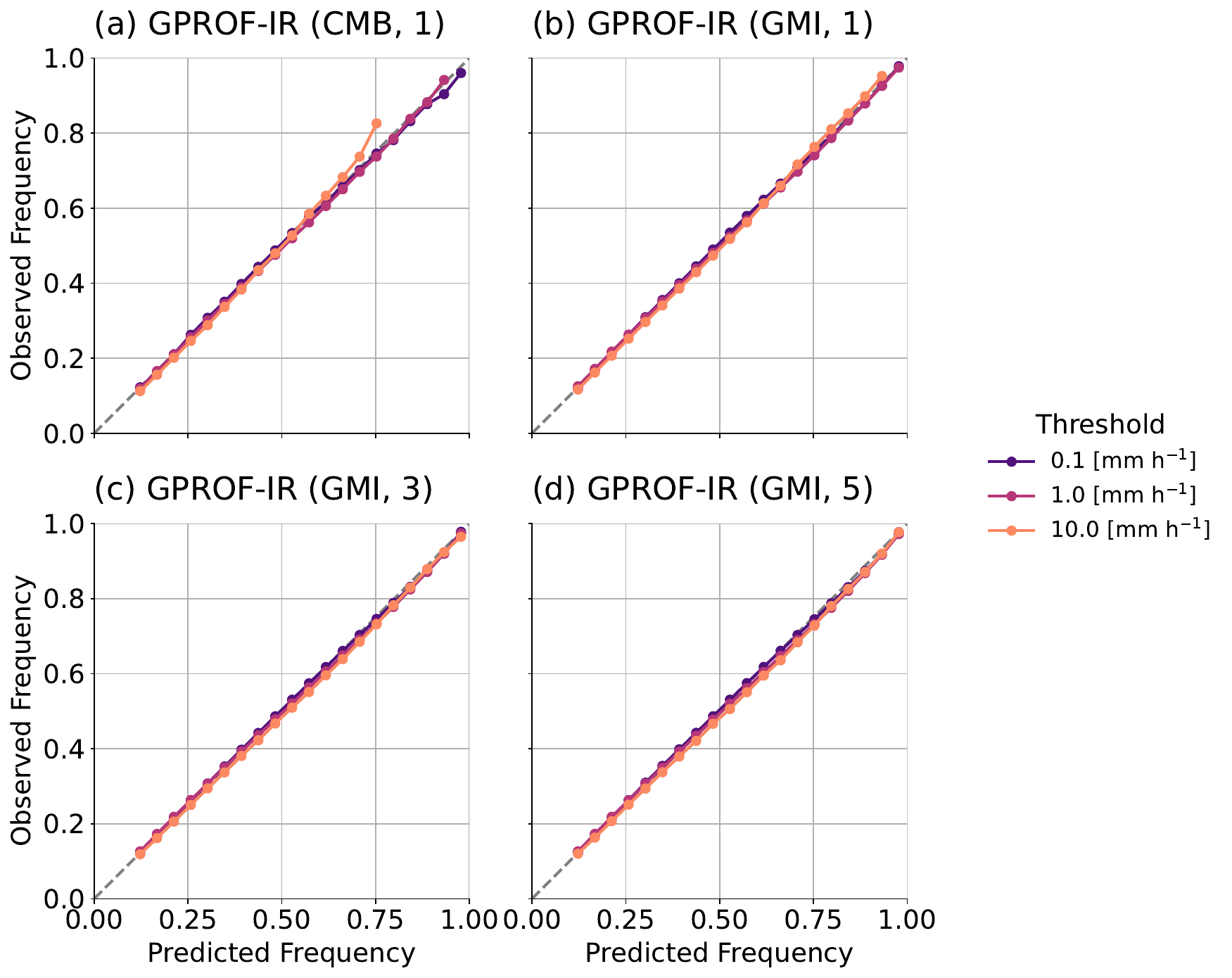}}
 \caption{
   Calibration of exceedance probabilities calculated using the
   probabilistic GPROF-IR retrieval output. Panels (a) -- (d) show the results for the
   four retrieval configurations. Line colors show the results for the three assessed
   thresholds (0.1 mm h$^{-1}$, 1.0 mm h$^{-1}$, 10.0 mm h$^{-1}$).
 }\label{fig:test_results_calibration}
\end{figure}

The probabilistic formulation of GPROF-IR provides information beyond the
posterior mean and enables an assessment of retrieval uncertainty. To evaluate
the calibration of the predicted posterior distributions, Fig.~\ref{fig:test_results_calibration} shows
reliability diagrams for exceedance probabilities at precipitation thresholds of
0.1 mm h$^{-1}$, 1.0 mm h$^{-1}$, and 10 mm h$^{-1}$. Across all thresholds, the
predicted exceedance probabilities  agree with the corresponding observed
frequencies, indicating that the retrievals are generally well calibrated and
provide reliable estimates of precipitation uncertainty.

\subsection{Independent Validation}

Because IMERG is designed to provide global precipitation estimates and is used
to study a wide range of meteorological processes, it is crucial to validate
the accuracy of GPROF-IR across diverse geographical regions and precipitation
regimes. Since no single reference dataset provides sufficiently accurate
precipitation estimates across all surfaces, regions, temporal scales, and
precipitation regimes, we combine complementary validation datasets to
assess GPROF-IR along three dimensions: quasi-instantaneous retrieval accuracy across
geographically distinct land regions, retrieval performance over ocean, and the
accuracy of temporally accumulated precipitation.

We use the SatRain testing dataset to assess instantaneous retrieval accuracy
over land because it provides independent reference precipitation estimates from
three geographically and climatologically distinct domains. Its public
availability makes the evaluation fully reproducible and facilitates future
comparisons with new retrieval algorithms. Because the SatRain reference data
are largely limited to land surfaces, we complement this evaluation with
collocations from the OceanRAIN dataset, which provides a rare opportunity to
validate precipitation estimates over the global oceans. Finally, we evaluate
hourly and daily precipitation accumulations over CONUS using temporally
continuous reference data from MRMS. This analysis assesses the ability of
GPROF-IR to reproduce precipitation over the lifetime of precipitation events
and enables a more detailed investigation of retrieval performance across
different precipitation regimes and over complex terrain.


\subsubsection{SatRain}

\begin{figure}[htbp]
 \centerline{\includegraphics[width=37pc]{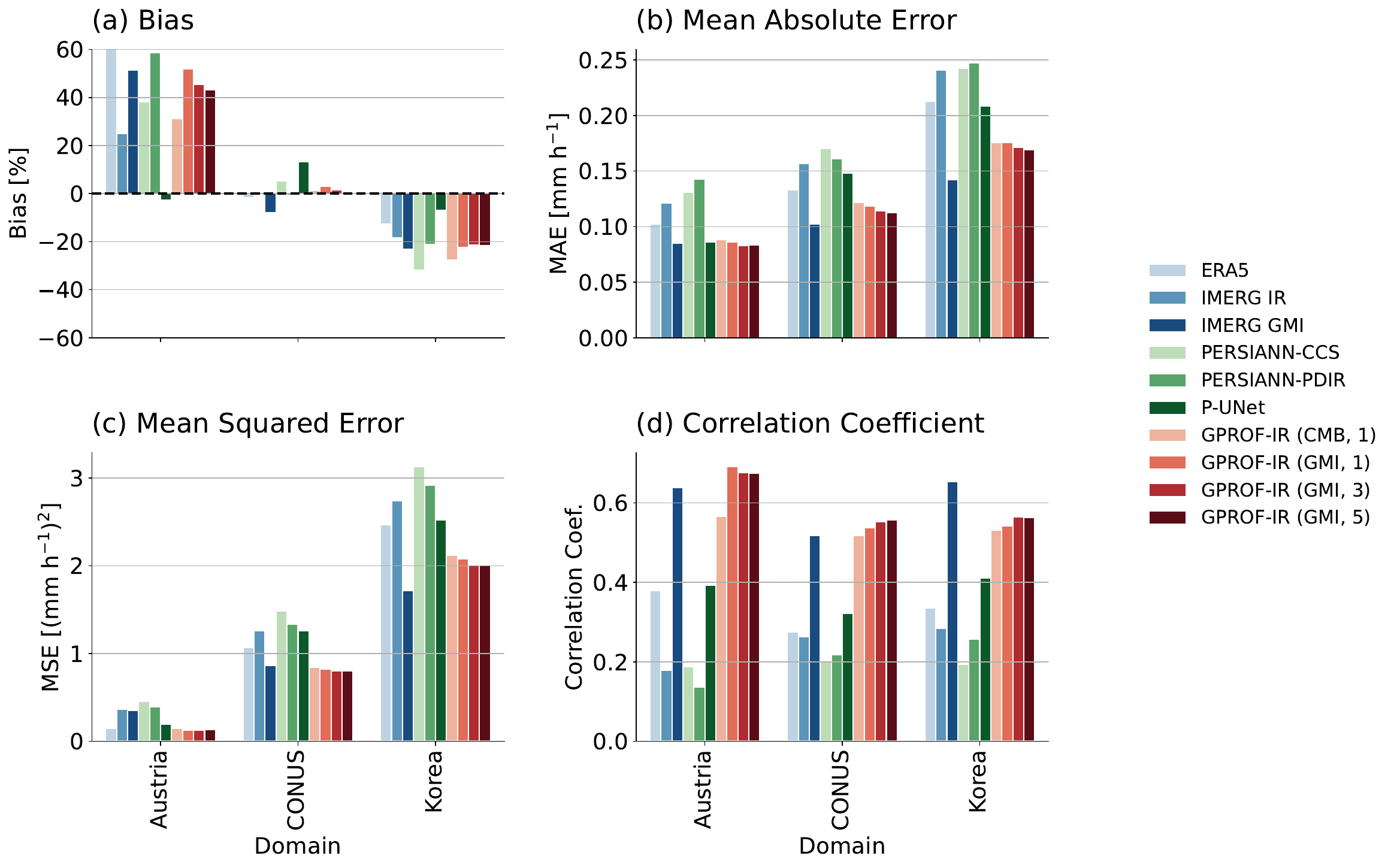}}
 \caption{
GPROF-IR retrieval accuracy assessed against baseline retrievals using the
SatRain benchmark dataset. Panels (a) – (d) compare bias, mean absolute error,
mean squared error, and correlation coefficient of the four GPROF-IR
configurations against ERA5, IMERG, and PERSIANN baselines. Each panel shows
results from the CONUS, Korea, and Austria testing domains of the SatRain
benchmark dataset.
 }\label{fig:results_satrain}
\end{figure}

Figure~\ref{fig:results_satrain} shows the evaluation metrics of the GPROF-IR
retrievals and the IMERG, ERA5, and PERSIANN baselines assessed across the three
testing domains of the SatRain dataset. In terms of biases, all baselines and
models exhibit similar tendencies: Strong positive biases over the Austria
domain, weak positive to no biases over CONUS, and moderate negative biases over
Korea. The regional differences in the biases are larger than the differences
between the different datasets. This indicates that they are mostly driven by
either the limited capability of remotely sensed and reanalysis precipitation
estimates to capture regional precipitation variability or inconsistencies in
the reference estimates.

In terms of mean absolute error, mean squared error, and correlation
coefficient, the GMI-based GPROF-IR (GMI, 1) retrieval performs consistently as
good or better than the 2BCMB-based GPROF-IR (CMB, 1) retrieval, indicating that
either the increased training data size or the improved representation of light
precipitation and precipitation over snow-covered surfaces of the GPROF V08
reference data increases the retrieval accuracy. In terms of the
incorporation of temporal information, the validation results confirm the
findings from the testing data: While the use of three timesteps yields a robust
improvement in retrieval accuracy, ingesting five instead of three timesteps
yields only marginal additional benefits.

The GPROF-IR retrieval greatly improves MAE, MSE, and linear correlation
compared to the IMERG V07 IR precipitation estimates despite using the same
satellite observations. Over the Austria and CONUS domains, the GPROF-IR
algorithms achieve better MSE and correlation coefficient than the IMERG GMI
precipitation estimates. Since GMI is among the most capable sensors of the GPM
constellation, the GPROF-IR retrievals  thus exceed the best-case accuracy of
IMERG V07 over these domains. It should be noted, however, that the PMW
retrievals in IMERG V07 use the Bayesian approach instead of the novel GPROF-NN
framework, which will be used in IMERG V08. For the Korea domain, the GPROF-IR
accuracy stays below that of the IMERG PMW estimates. This is likely because the
data from the Korea domain contains relatively more precipitation estimates over
water surfaces, where most PMW sensors have higher information content.
Additionally, warm precipitation with lower cloud tops and thus weaker
signatures in the IR imagery plays a larger role over the Korea domain than over
CONUS, which may further contribute to the relatively lower accuracy of the IR
retrievals compared to the IMERG GMI estimates.

The GPROF-IR retrievals also yield substantially higher accuracy than ERA5 and
PERSIANN at the quasi-instantaneous timescales assessed by the SatRain dataset.
Both ERA5 and PERSIANN provide hourly precipitation estimates, whereas the
SatRain reference measurements are quasi-instantaneous. This mismatch in
temporal support introduces additional representativeness error and places these
hourly products at a disadvantage in this comparison. ERA5 and PERSIANN are
included here as contextual baselines rather than to assess their performance at
their native temporal resolution. It is nonetheless notable that the IMERG IR
precipitation estimates consistently yield higher errors and lower correlation
than the ERA5 precipitation, indicating that its higher spatial and temporal
resolution does not improve the representation of precipitation compared to the
coarser ERA5 estimates.

\paragraph{Effective Resolution}

The effective resolution provides a measure of the smallest scale of
precipitation features that a retrieval is able to capture. Since the analysis
of the effective resolution requires spatially continuous precipitation fields
covering at least the resolved scales, the SatRain dataset only allows assessing
the effective resolution over the CONUS and Korea domains.

\begin{figure}[htbp]
 \centerline{\includegraphics[width=37pc]{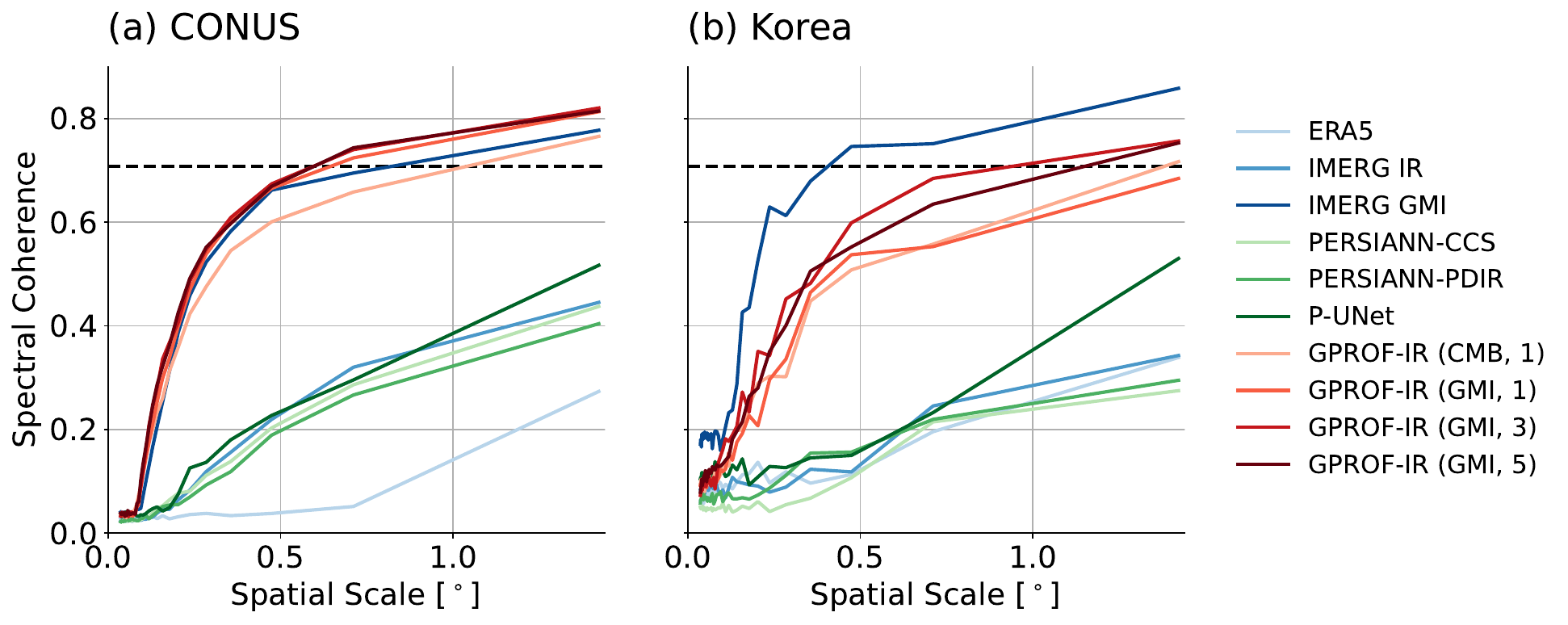}}
 \caption{
  Spectral coherence curves for the GPROF-IR retrievals and baseline
precipitation estimates assessed using the CONUS and Korea domains of the
SatRain benchmark dataset. Panels (a) and (b) show the spectral coherence across
scales in degree for the CONUS and Korea domains, respectively. The dashed black
line marks the $\sqrt{\frac{1}{2}}$ threshold used to define the effective
resolution of the precipitation estimates.
 }\label{fig:spectral_coherence}
\end{figure}

The spectral coherence curves for the CONUS and Korea domains are shown in
Fig.~\ref{fig:spectral_coherence}. Over CONUS, the GPROF-IR (GMI, 3) and GPROF-IR
(GMI, 5) configurations achieve effective resolutions of 0.6$^\circ$ whereas the
GPROF-IR (GMI, 1) retrieval achieves an effective resolution of 0.63. The 2BCMB-based
retrieval only achieves an effective resolution of 0.9$^\circ$. Over the Korea
domain, the effective resolutions are lower with resolutions around 0.84$^\circ$
for  GPROF-IR (GMI, 3) and 1.07$^\circ$ for the GPROF-IR (GMI, 5) retrieval
whereas and the GPROF-IR (GMI, 1) and GPROF-IR (CMB) configurations do not reach
the $\sqrt{\frac{1}{2}}$ threshold within the resolved scale range.

 The GPROF-IR (GMI, 3) and GPROF-IR (GMI, 5) retrievals achieve effective
resolutions that are about 25\% finer than the GMI-based IMERG V07 GMI estimates
over CONUS. However, the situation is reversed over the Korea domain, where the
IMERG V07 GMI precipitation estimates are about 33 \% finer than the best
GPROF-IR retrievals. The IMERG IR, ERA5, and PERSIANN precipitation fields stay
below the $\sqrt{\frac{1}{2}}$ threshold throughout the resolved scale range indicating
that they can't reliably resolve precipitation structures finer than
$1.4^\circ$.

\paragraph{Precipitation Detection}

\begin{figure}[htbp]
 \centerline{\includegraphics[width=37pc]{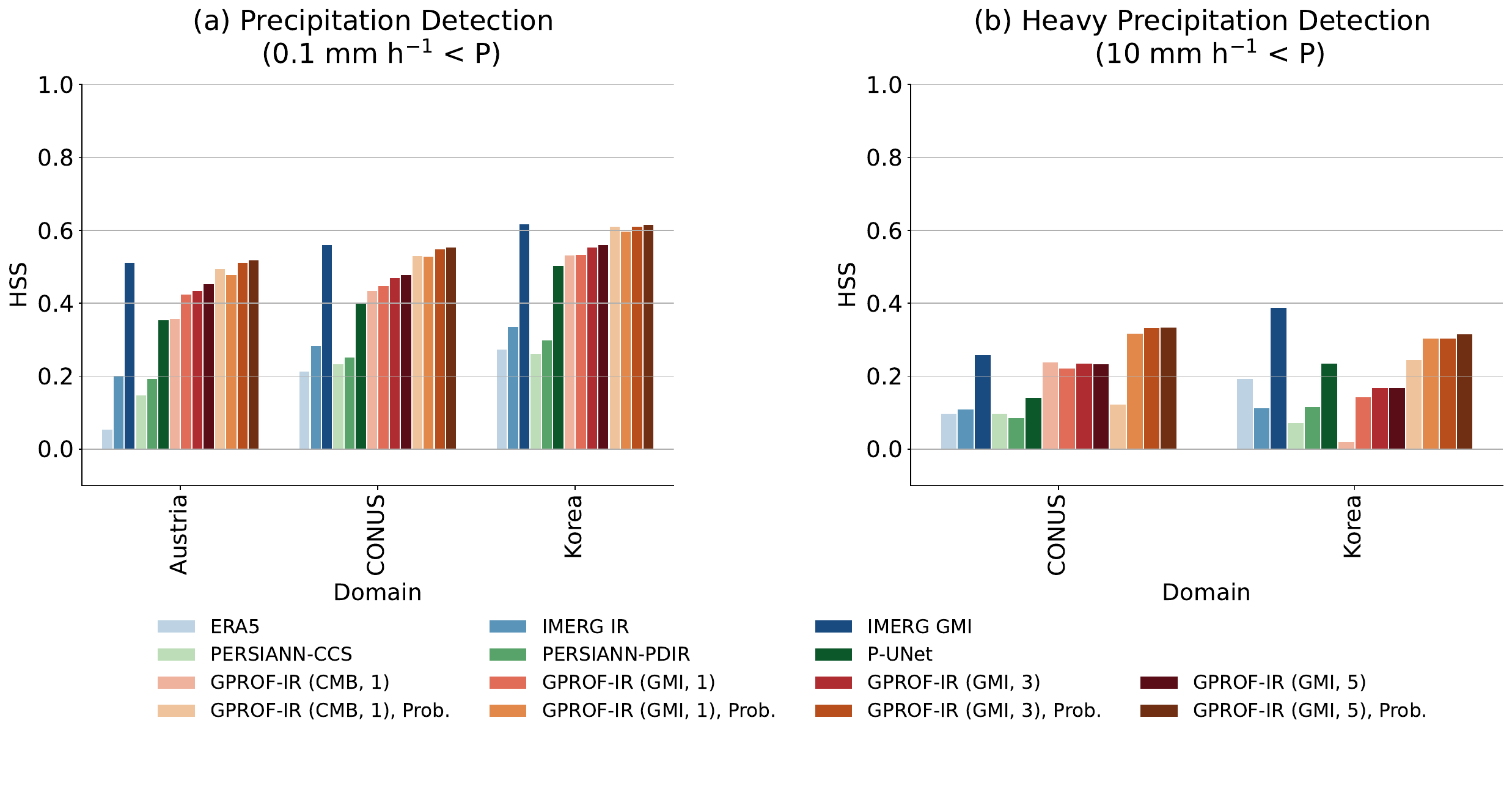}}
 \caption{
   Detection skill for precipitation and heavy precipitation assessed on the
SatRain benchmark dataset. Panel (a) shows the Heidke Skill Score (HSS) for the
detection of precipitation (defined as any grid point exceeding $0.1$ mm
h$^{-1}$). Panel (b) shows the corresponding results for the detection of heavy
precipitation (defined as grid points exceeding 10 mm h$^{-1}$). Results for
GPROF-IR are presented using the retrieved precipitation rate as detection
criterion (red shades) and using the probabilistic output to calculate
exceedance probabilities with optimal detection thresholds calculated using the
validation dataset.
 }\label{fig:precipitation_detection}
\end{figure}

Figure~\ref{fig:precipitation_detection} assesses the skill of the GPROF-IR
retrievals to detect precipitation and heavy precipitation identified using
thresholds of 0.1 mm h$^{-1}$ and 10 mm h$^{-1}$, respectively. For the GPROF-IR
retrievals, two different detection approaches are compared: The first approach
(red shading) is to use the precipitation rate corresponding to the mean of the
posterior distribution, which is provided as the principal quantitative
precipitation output from the retrieval. The second approach is to use the
probabilistic output of the model to calculate the probability of the observed
precipitation rate to exceed the respective threshold (orange shading). Optimal
detection thresholds for light and heavy precipitation and each retrieval
configuration were determined on the validation dataset.

Using the retrieved GPROF-IR precipitation rates for precipitation detection
yields higher skill than the IMERG IR and ERA5 baselines, but lower skill than
the IMERG GMI estimates. Detection performance improves when exceedance
probabilities are used instead of the retrieved precipitation rates. With the
probabilistic output, the multi-timestep GPROF-IR (GMI, 3) and GPROF-IR (GMI, 5)
retrievals achieve detection skill comparable to that of the IMERG GMI
estimates.

Detection of heavy precipitation is more challenging, with substantially lower
HSS values for all retrievals. Over CONUS, detection based on the GPROF-IR
precipitation rates remains less skillful than IMERG GMI, whereas use of the
probabilistic GPROF-IR output increases the detection skill beyond that of the
IMERG GMI estimates. Over Korea, the probabilistic output likewise improves
heavy-precipitation detection, although the resulting skill remains below that
of IMERG V07.

 \paragraph{Case Studies}

We further assess the accuracy of the GPROF-IR retrievals using two case studies
over the Korea and CONUS domains.

\begin{figure}[htbp]
 \centerline{\includegraphics[width=37pc]{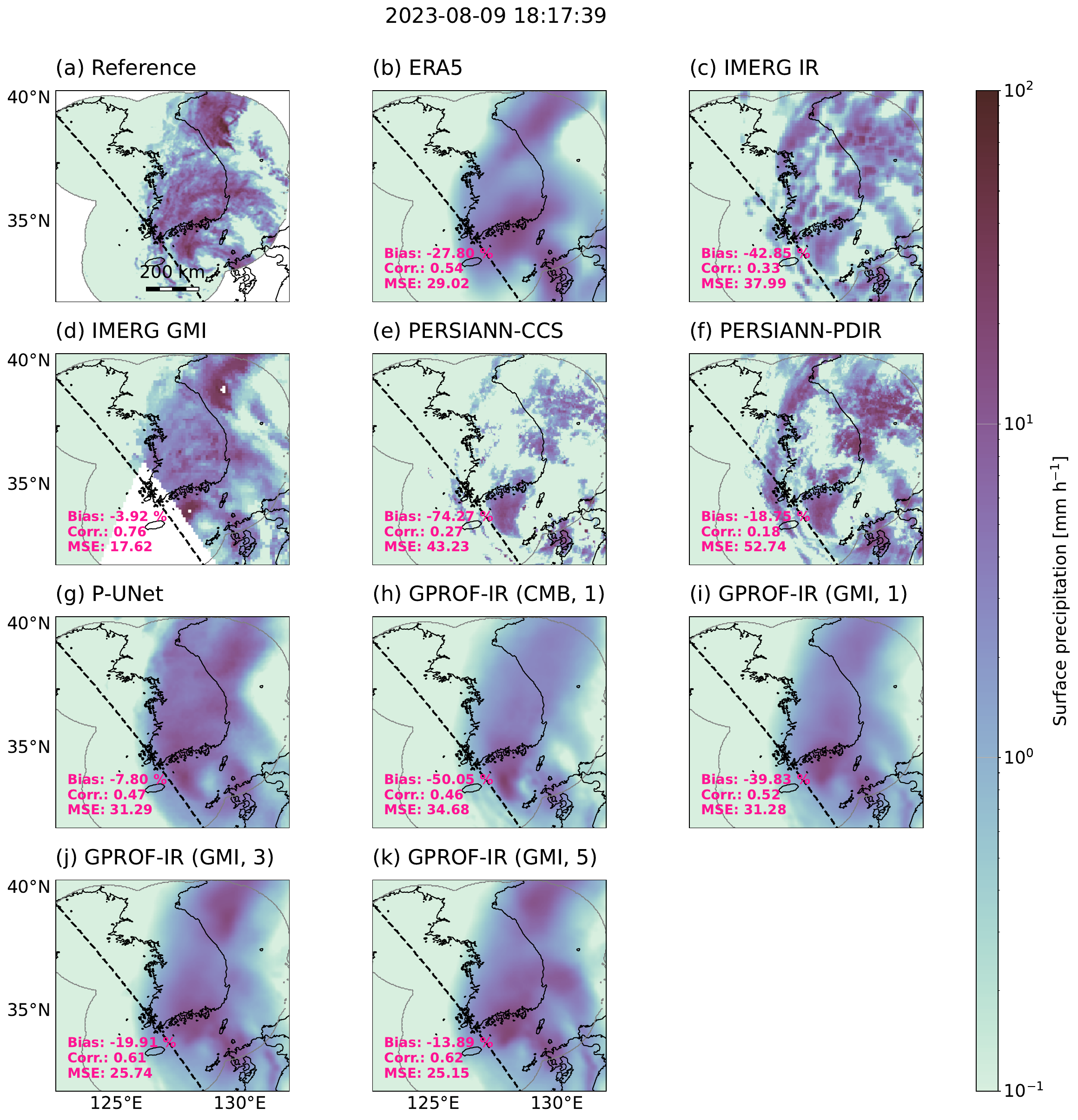}}
 \caption{
    Reference and retrieved precipitation during landfall of Typhoon Khanun on the
    Korean peninsula. Panel (a) shows reference precipitation estimates from
    gauge-corrected ground-based radars from the SatRain benchmark dataset. Panels
    (b) – (h) show the results from the ERA5 and IMERG baselines as well as the
    GPROF-IR retrievals.
 }\label{fig:case_study_1}
\end{figure}

Figure~\ref{fig:case_study_1} shows reference and retrieved precipitation during
the landfall of Typhoon Khanun over the Korean peninsula \citep{bi_khanun_2025}.
The principal precipitation features in the reference precipitation field are
two strong cyclonic rain bands in the south-eastern quadrant of the domain as
well as strong convection in the north-eastern quadrant. The ERA5 estimates
capture the general cyclonic structure and the precipitation in the
north-eastern part of the scene but lack finer convective structures. The IMERG
GMI precipitation estimates yield the best agreement with the reference
estimates capturing the principal rain bands as well as the embedded convection.
The IMERG IR precipitation estimates fail to reproduce the structure of the
cyclone. Although the estimates contain substantial small-scale variability, it
is unrelated to any of the fine-scale structures in the reference data.

Despite corresponding to hourly accumulations, the PERSIANN CCS and PDIR-Now
precipitation estimates exhibit even finer precipitation structures than the
IMERG IR precipitation fields. However, they similarly fail to capture the large
scale precipitation features resulting in high mean-squared error and low linear
correlations. The P-UNet retrieval produces a smoother precipitation field that
better captures the large-scale precipitation features and achieves lower bias as
well as a lower MSE and higher correlation coefficient.

Similarly, the GPROF-IR retrievals capture the large-scale precipitation but
lack the fine precipitation features visible in the radar measurements. The
inclusion of temporal information yields notable and visible improvements with
the retrieval being able to reproduce more of the embedded precipitation
features and reducing the negative bias from -38 \% for the GPROF-IR (CMB, 1)
retrieval to -14 \% for the GPROF-IR (GMI, 5) retrieval.

\begin{figure}[htbp]
 \centerline{\includegraphics[width=37pc]{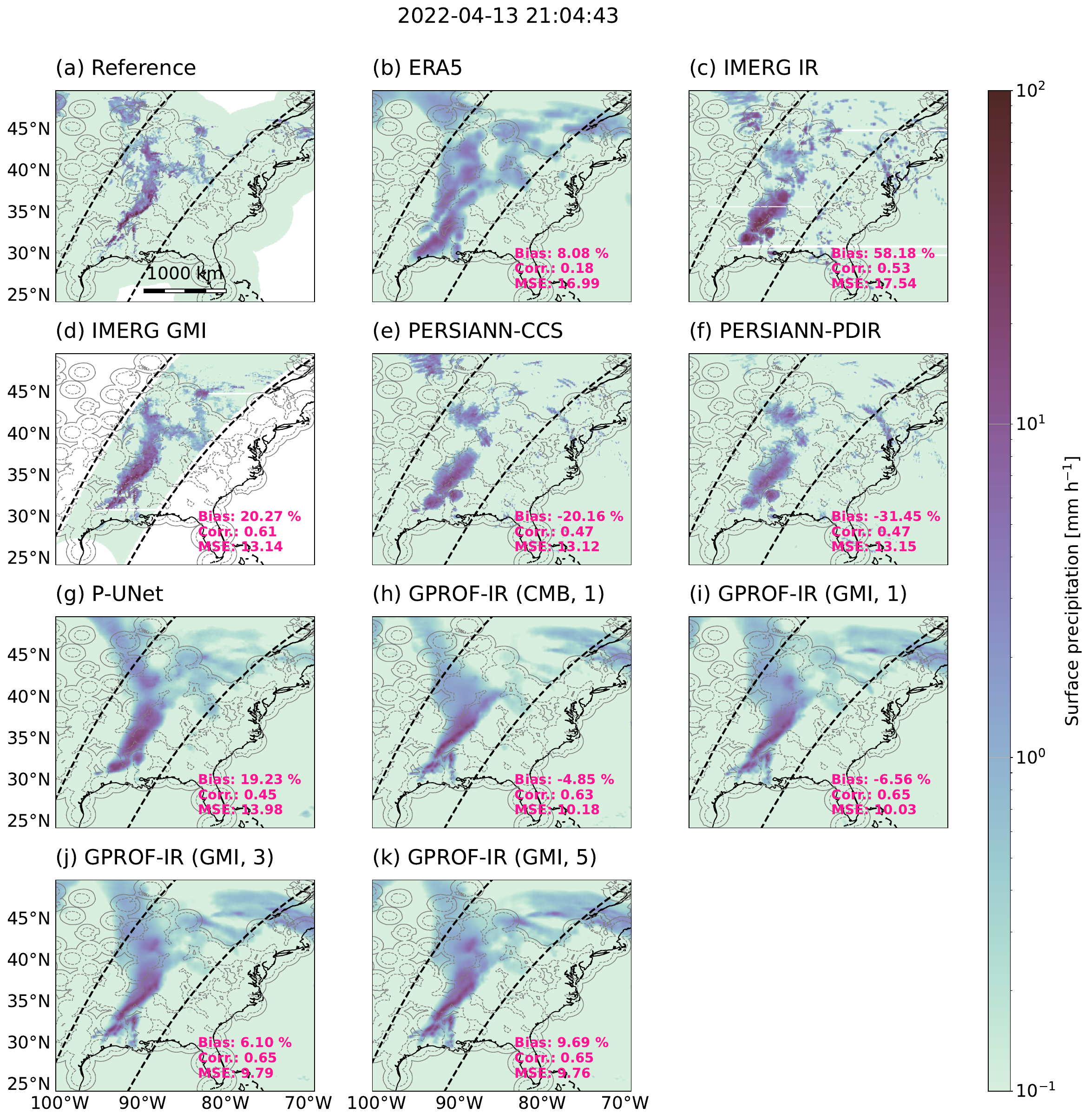}}
 \caption{
    Reference and retrieved precipitation for a convective storm system on
    April 13, 2022 over CONUS. Panel (a) shows reference precipitation estimates
    from gauge-corrected ground-based radars from the SatRain benchmark dataset.
    Panels (b) – (k) show the results from the ERA5, IMERG, and PERSIANN baselines
    as well as the GPROF-IR retrievals.
 }\label{fig:case_study_2}
\end{figure}

Figure~\ref{fig:case_study_2} shows a second case study  of a storm system
over CONUS that spawned several Tornadoes on April 13, 2022. The principal
precipitation features are a convective squall line in the south-western
quadrant and lighter, scattered precipitation in the north. For this case, the
ERA5 estimates show the worst agreement with the reference data. Although ERA5
captures some of the lighter precipitation in the North, it misplaces the
convection in the South and exaggerates its spatial extent. The IMERG GMI
estimates yield good agreement with the reference data capturing both the squall
line and the lighter precipitation. The IMERG IR estimates capture some of the
heavy convective precipitation but fail to reproduce the spatial organization of
the squall line. Furthermore, the light precipitation observed in the North
bears little resemblance with the reference data.

The PERSIANN CCS and PDIR retrievals produce estimates that are close to those of the
IMERG IR fields. The P-UNet retrieval captures more precipitation in the northern
part of the scene but overestimates the spatial extent of the precipitation
leading to higher error than for the CCS and PDIR retrievals.

The GPROF-IR retrievals yield the best accuracy for this case study, accurately
capturing the squall line albeit slightly exaggerating its spatial extent. All
configurations slightly overestimate the spatial coverage of precipitation in
the North but generally capture all of the important precipitation features
visible in the reference data. The GPROF-IR retrieval likely performs well for
this case study because IR brightness temperatures are more directly linked to
intense precipitation on the ground. Combined with the reduced information
content in the PMW observations due to background emission from the land
surface, the GPROF-IR retrievals are able to produce more accurate estimates
than the GMI-based IMERG estimates.

\subsubsection{OceanRAIN}

The validation using the SatRain dataset was largely limited to measurements
over land. We complement the land-focused validation with an evaluation of all
retrievals against shipborne in-situ precipitation measurements from the
OceanRAIN dataset. The GPROF V08 training data used to train the GMI-based
GPROF-IR retrievals uses the GPROF-NN XPR retrieval to fill in missing light
precipitation in the 2BCMB dataset over oceans. This has been shown to reduce
systematic underestimation of high-latitude and frozen precipitation
\citep{pfreundschuh_xpr_2026}. However, the analysis was limited to GMI
overpasses over the OceanRAIN vessels. The continuous availability of the IR
observations thus provides an additional opportunity to indirectly validate the
representation of light and frozen precipitation in GPROF V08 retrievals using a
 larger number of samples.

\begin{figure}[htbp]
 \centerline{\includegraphics[width=37pc]{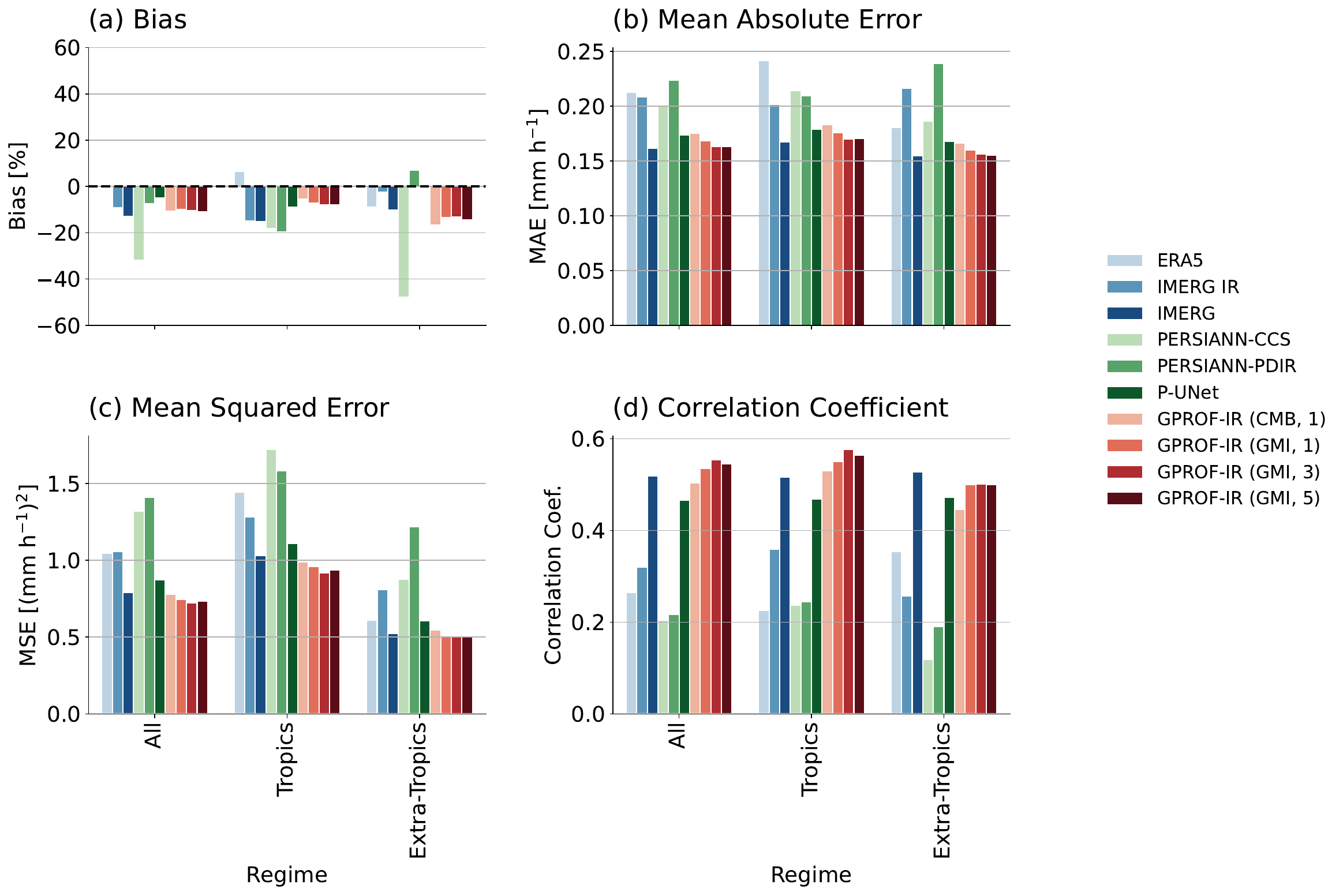}}
 \caption{
   Validation of precipitation estimates over ocean using two years (2017 and
2018) of shipborne distrometer measurements from the OceanRAIN dataset. Panels
(a) – (d) compare bias, mean absolute error, mean squared error, and correlation
coefficient of the two of GPROF-IR configurations against ERA5, IMERG, and
PERSIANN baseline estimates. Each panel shows results for all matchups,
tropics (equatorward of 30$^\circ$) and extratropics (poleward of 30$^{\circ}$)
 }\label{fig:ocean_rain_results}
\end{figure}

The accuracy metrics of all retrievals are shown in Fig.~\ref{fig:ocean_rain_results}.
With the exception of ERA5, all datasets exhibit a tendency to underestimate
total oceanic precipitation. While most datasets exhibit small biases of between
-5 to -10 \%, the PERSIANN CCS retrieval underestimates total precipitation by
-32 \%. Most retrievals exhibit similar bias tendencies in tropics as the
extratropics with the exception of ERA5, which slightly overestimates
precipitation in the tropics but underestimates precipitation the extra-tropics,
and PERSIANN PDIR, which underestimates precipitation in the tropics but
overestimates it in the extra-tropics.

In terms of retrieval errors, the multi-step GPROF-IR retrievals yield the most
accurate retrievals overall. GPROF-IR (GMI, 3) and GPROF-IR (GMI, 5) achieve
slightly higher accuracy than the merged IMERG V07 estimates with larger
improvements in the tropics but slightly lower accuracy in the extra-tropics. It
should be noted that the IMERG estimates here correspond to the merged estimates
combining both the PMW and IR estimates since -- in contrast to the SatRain data
-- the validation data is not limited to GMI overpasses. Moreover, the analysis
explicitly excludes GMI overpasses, which may slightly underestimate the
accuracy of the merged IMERG V07 estimates. The P-UNet retrieval also achieves
comparably high accuracy but remains below the accuracy of the merged IMERG
estimates. The accuracy of the ERA5, IMERG-IR, PERSIANN CCS, and PERSIANN PDIR
baselines remains considerably lower than the merged IMERG, GPROF-IR, or P-UNet
estimates.

\subsubsection{Hourly and Daily Accumulations over CONUS}

Finally, we assess the ability of the GPROF-IR retrievals to reproduce hourly
and daily precipitation accumulations by comparing retrieved accumulations
against gauge-corrected reference measurements from ground based radars. This
analysis complements the SatRain-based evaluation of quasi-instantaneous
precipitation by assessing retrieval performance continuously over a full year
and at longer temporal scales. The dense temporal sampling also enables the
seasonal and spatial variability of the retrieval errors to be assessed,
providing a more comprehensive evaluation across the precipitation regimes
encountered over CONUS.

\begin{figure}[htbp]
 \centerline{\includegraphics[width=37pc]{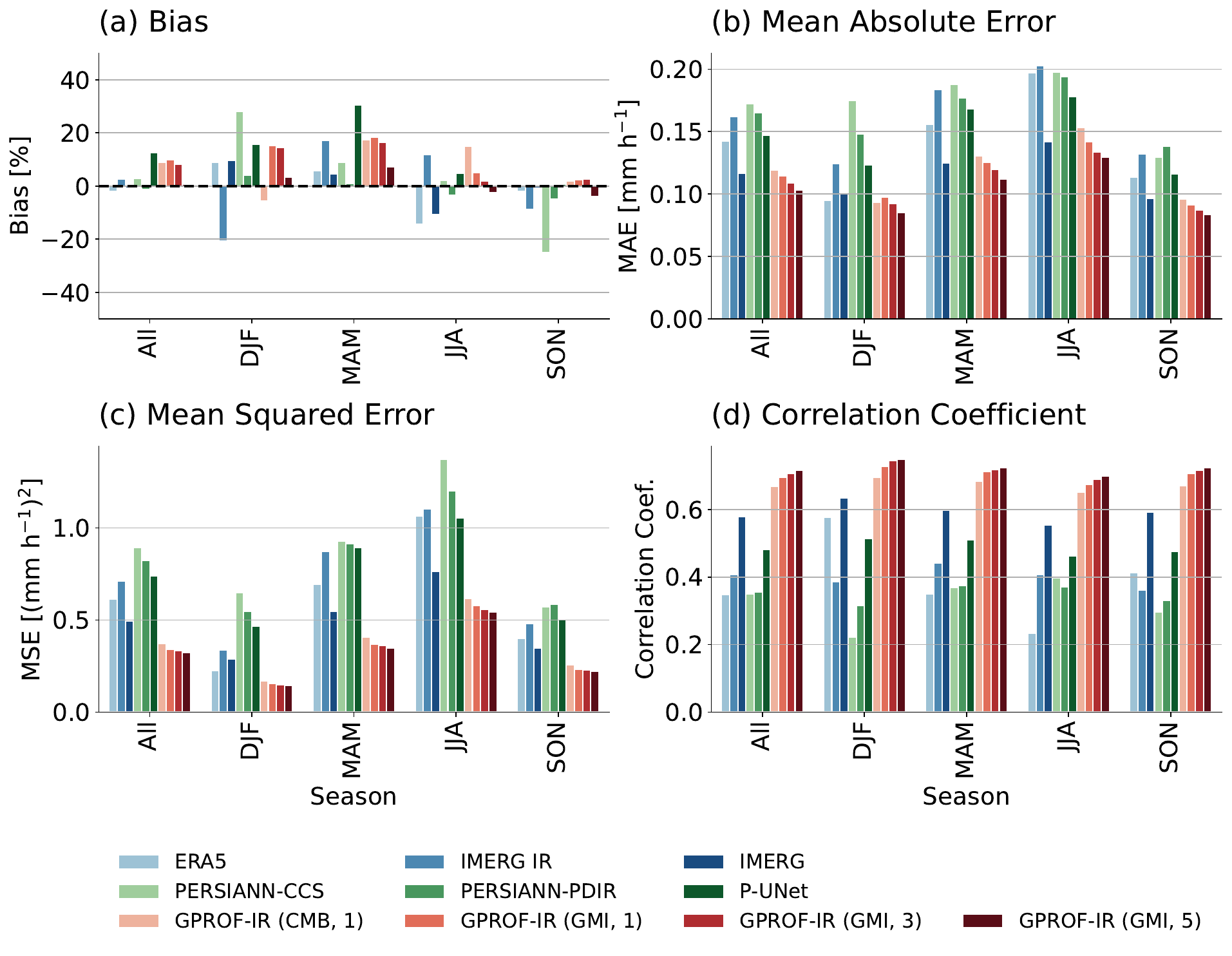}}
 \caption{
   Accuracy metrics of hourly precipitation accumulations for the year 2022
compared to gauge-corrected radar estimates from MRMS. Panels
 show the results for (a) the bias, (b) mean absolute error (MAE), (c) mean squared error
(MSE) and correlation coefficient.
 }\label{fig:results_conus_1h}
\end{figure}

The evaluation of the hourly accumulations shown in
Fig.~\ref{fig:results_conus_1h}. All retrievals achieve improved error
statistics at the hourly scale, as temporal accumulation reduces random
retrieval errors and spatial variability in the reference data are reduced, but
the results remain similar the validation using quasi-instantaneous SatRain test
data. The comparison against hourly accumulation also reduces representativeness
errors affecting the ERA5, PERSIANN-CCS, PERSIANN-PDIR, and P-UNet estimates,
which are only available as hourly accumulations. The relative improvement in
the hourly accumulations remains smaller for IMERG since this analysis
necessarily includes estimates from other sensors than GMI. The difference in
sampling leads to a reduction of the accuracy of IMERG relative the other
datasets. Among the satellite retrievals, IMERG IR, PERSIANN-CCS and
PERSIANN-PDIR still exhibit the largest errors. P-UNet improves upon the
conventional IR-based precipitation estimates but remains less accurate than
IMERG. The GPROF-IR retrievals provide the most accurate estimates among the
evaluated IR-based retrievals while also outperforming IMERG, yielding 33 \%
lower MSE and 20 \% higher correlations over the full year.

The annual-mean biases of ERA5, IMERG, IMERG IR, and PERSIANN PDIR are close to
zero, whereas PERSIANN CCS, P-UNet, and the GPROF-IR retrievals exhibit small
positive biases that occur primarily during winter and spring. Although, MAE and
MSE exhibit a pronounced seasonal cycle for all retrievals, with larger absolute
errors during the summer months when heavier convective precipitation becomes
more prevalent, the relative accuracy of the satellite precipitation retrievals
is stable across the seasons. ERA5, on the other hand, shows substantially
stronger seasonal variability in its performance relative to the satellite-based
estimates. During winter, ERA5 yields MAE and MSE values comparable to those of
IMERG, whereas its relative performance deteriorates through spring and reaches
a minimum during summer. This seasonal dependence is consistent with greater
ERA5 skill for large-scale, synoptically driven precipitation and lower accuracy
for warm-season convective precipitation.

\begin{figure}[htbp]
 \centerline{\includegraphics[width=37pc]{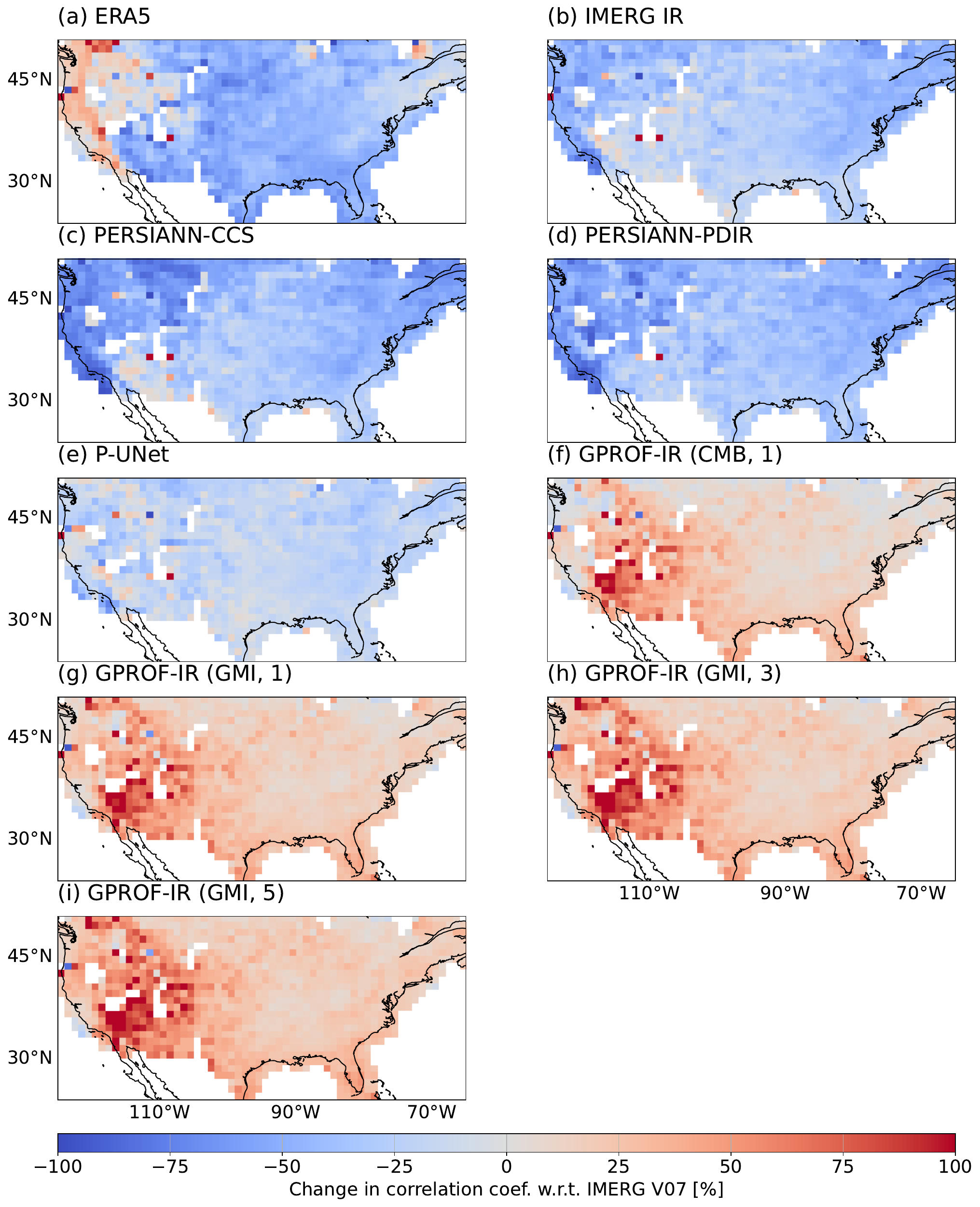}}
 \caption{
   Relative difference in the correlation coefficient compared to IMERG V07
   for hourly precipitation accumulations compared to gauge-corrected ground-based
   radar estimates from MRMS.
 }\label{fig:improvements_conus}
\end{figure}

Since the climate as well as dominant precipitation regimes vary substantially across
CONUS, the spatial dependence of retrieval accuracy provides an opportunity to
assess the robustness of the retrievals across different precipitation
environments.  Fig.~\ref{fig:improvements_conus} shows the spatial
distribution of the difference in the correlation coefficient between all
assessed datasets and IMERG V07. The conventional IR algorithms (IMERG IR,
PERSIANN-CCS, and PERSIANN-PDIR) exhibit lower correlations than IMERG V07
across nearly all of CONUS, with  little spatial variability in
their performance relative to IMERG. ERA5, in contrast, achieves higher
correlations than IMERG in the western mountain ranges. This pattern is
consistent with improved ERA5 skill for the large-scale,
orographically influenced precipitation that is common in these regions.
However, the improvement is not apparent across all mountainous regions and may
also partly reflect regional variations in the accuracy of IMERG.

The GPROF-IR retrievals yield higher correlation than IMERG V07 across nearly
all of CONUS, with the largest improvements occurring in the south-western
United States. Comparison with the absolute correlation coefficients shown in
Fig.~\ref{fig:conus_spatial_corr} reveals that these enhanced improvements
coincide with regions in which IMERG V07 exhibits lower correlations. In
contrast, the absolute correlation values for the GPROF-IR retrievals vary less
strongly across CONUS and do not exhibit the same degradation over much of the
Mountain West as IMERG. These results indicate that GPROF-IR provides more
spatially consistent retrieval performance across the diverse precipitation
environments encountered over CONUS, including regions in which IMERG V07
exhibits reduced skill.

At daily scales, the accuracy of all retrievals increases further. The GPROF-IR
estimates remain the most accurate overall although the IMERG V07 achieves
values closer to those of GPROF-IR (CMB, 1), the least accurate GPROF-IR
configurations. While the overall accuracy of ERA5 remains lower than that of
IMERG V07 or the GPROF-IR retrievals it slightly surpasses the accuracy of the
GPROF-IR retrievals during the winter months. However, the accuracy notably
decreases during spring and summer leading to lower overall accuracy. The large
performance increase of the ERA5 precipitation at these longer accumulation
times should be interpreted in light of its assimilation of 6-hourly Stage IV
precipitation estimates over parts of CONUS \citep{hersbach_era5_2020}. Because
CONUS is the only region where surface precipitation estimates are assimilated
in ERA5, it is not clear that ERA5 can provide equally accurate estimates
outside of CONUS.

The PERSIANN retrievals also improve but remain less accurate than ERA5, IMERG
V07, or the GPROF-IR retrievals. The exception is during summer when the
accuracy of P-UNet increases over that of ERA5.

\begin{figure}[htbp]
 \centerline{\includegraphics[width=37pc]{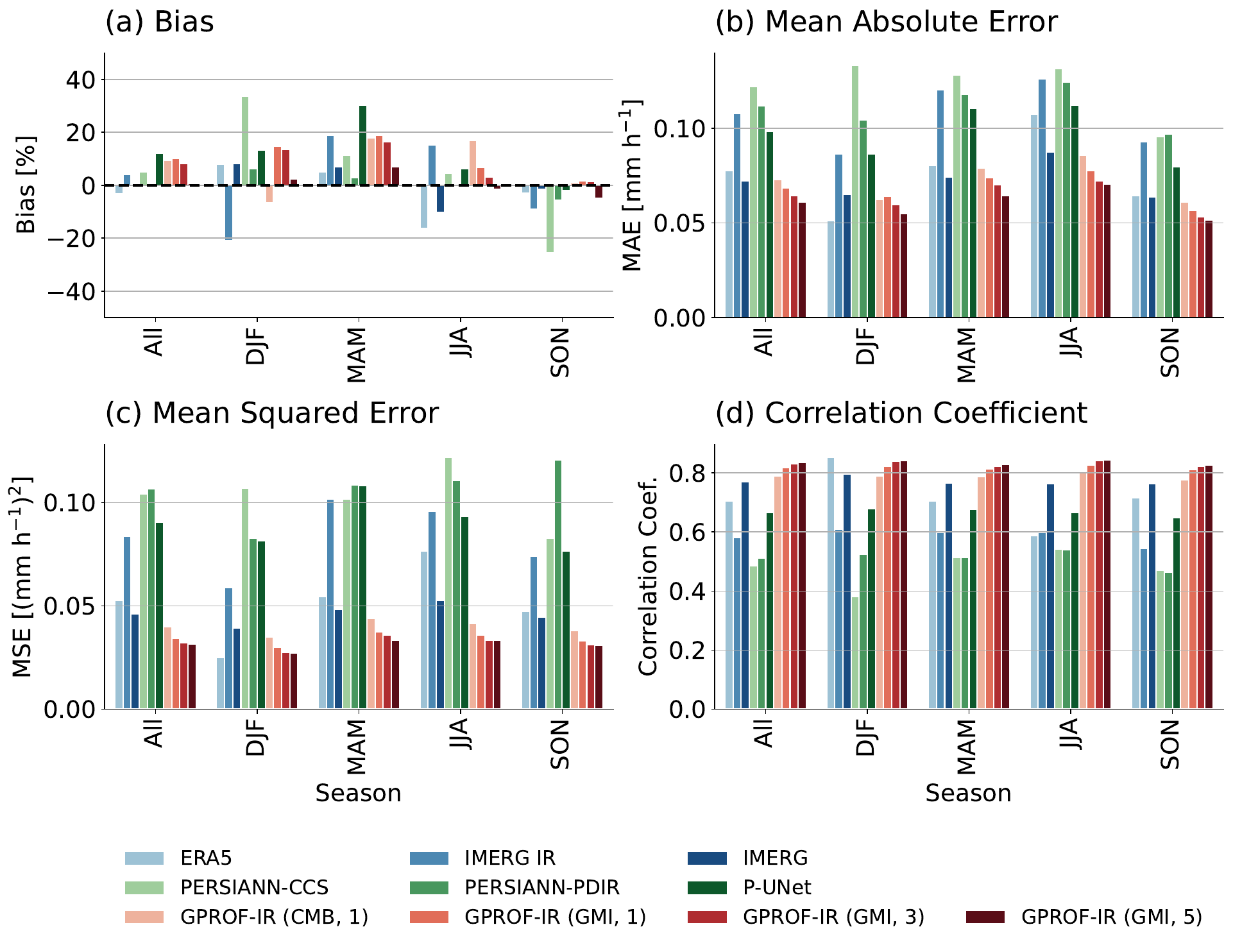}}
 \caption{
   As Fig.~\ref{fig:results_conus_1h} but for 24-hour accumulations.
 }\label{fig:results_conus_24h}
\end{figure}

\subsection{Precipitation Climatology}

Below we characterize the global precipitation climatologies of the GPROF-IR
(GMI, 3) configuration, which will be used operationally for IMERG V08. The
climatologies are derived using retrievals from the year 2022. To reduce
differences due to different spatiotemporal sampling, precipitation estimates
from GPROF-IR, ERA5, IMERG V07 and GPCP are extracted matching the sampling of
GMI observations during 2022.

\begin{figure}[htbp]
 \centerline{\includegraphics[width=37pc]{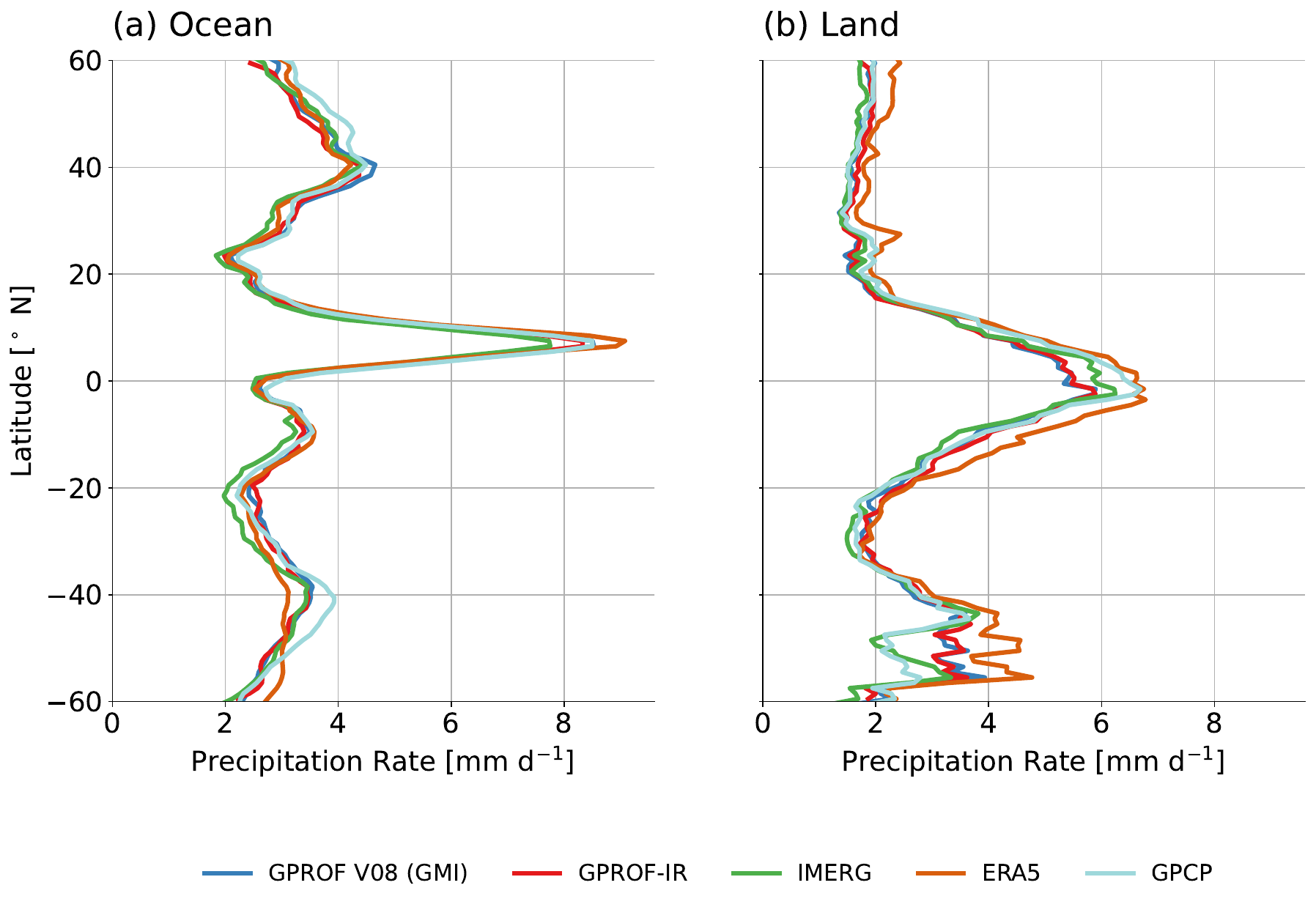}}
 \caption{
    Zonal mean precipitation from GPROF V08, GPROF-IR, IMERG, ERA5 and GPCP for
2022. Panel (a) displays the means for precipitation over oceans; panel (b) the
means for precipitation over land.
 }\label{fig:zonal_means}
\end{figure}

Figure~\ref{fig:zonal_means} shows zonal means from the GPROF-IR (GMI, 3)
retrieval compared to corresponding estimates from GPROF V08 GMI, IMERG V07,
ERA5 and GPCP. As expected, GPROF-IR results closely track the GPROF V08
reference estimates over both ocean and land. The datasets largely agree
over the tropical oceans with IMERG V07 yielding a slightly lower peak and ERA5 a slightly
higher peak in the intertropical convergence zone than GPCP, GPROF V08 GMI, and GPROF-IR.
Over northern-hemisphere oceans,  GPROF V08 GMI, GPROF-IR, IMERG V07 and ERA5  show
close agreement while the GPCP estimates remain higher. Over the southern
 ocean, the GPROF V08 GMI, GPROF-IR, and IMERG estimates lie between the ERA5
and GPCP estimates at around 40$^\circ$ S but fall below the ERA5 estimates
poleward of 50$^\circ$ S.

Over land, the discrepancies between the datasets are larger. Both GPCP and ERA5
are about 20\% higher than GPROF V08 GMI and GPROF-IR over tropical land. Over
northern hemisphere land, GPROF V08, GPROF-IR, IMERG V07, and GPCP show good
agreement while ERA5 yields slightly higher average precipitation rates. Over
southern-hemisphere land, the differences are larger with GPCP and IMERG V07
being substantially lower than GPROF V08 and GPROF-IR while ERA5 remains higher.
However, the land area contributing to the zonal means becomes small at these
latitudes, making the averages increasingly sensitive to differences over a
limited number of geographic regions.

\begin{figure}[htbp]
 \centerline{\includegraphics[width=37pc]{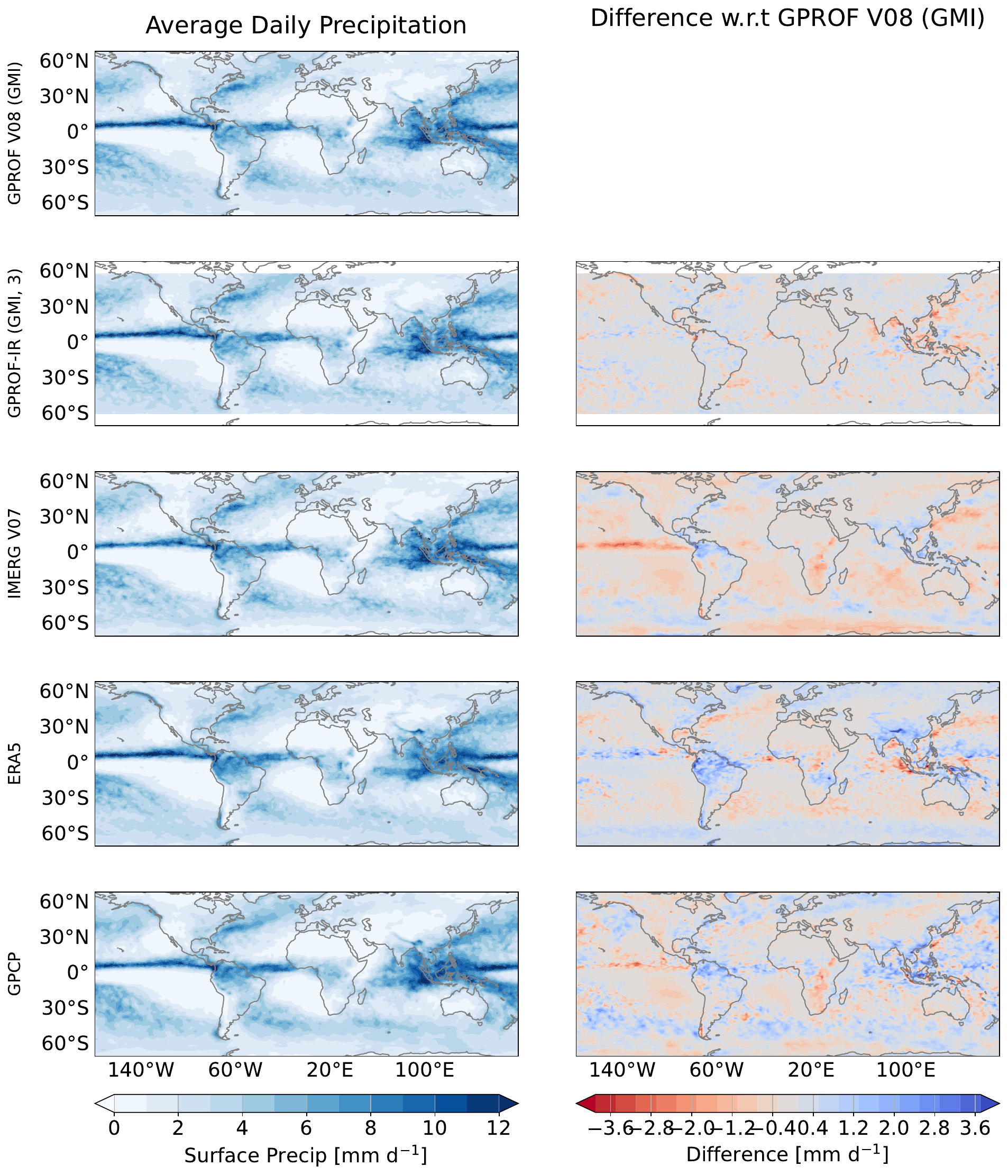}}
 \caption{
Global precipitation distributions from GPROF V08, GPROF-IR, ERA5, and GPCP for
2022. The first column displays the average precipitation distributions; the
second column displays the difference with respect to GPROF V08.
 }\label{fig:means}
\end{figure}

Figure~\ref{fig:means} shows the global precipitation distributions for GPROF
V08 for GMI, GPROF-IR, IMERG V07, ERA5, and GPCP. The distribution of the
GPROF-IR estimates closely resembles that of GPROF V08. Differences between
GPROF V08 and GPROF-IR are mostly irregularly distributed and don’t exhibit any
notable spatial patterns, indicating that they are largely driven by random
retrieval errors for individual precipitation events. Both ERA5 and GPCP show
larger systematic differences particularly over land. ERA5 tends to be higher
over land than both GPROF V08 and GPROF-IR, particularly in the tropics. For
GPCP the differences are less uniform with regions of both underestimation and
overestimation present at all latitudes.

The GPROF-IR (GMI, 3) retrievals largely reproduce the distribution of GPROF
V08. Since the GPROF-IR retrieval is designed to provide background
precipitation estimates to be merged with GPROF V08 retrievals, the
climatological consistency constitutes an additional desirable feature of the
GPROF-IR retrievals.

\section{Discussion}
This study developed and validated GPROF-IR — a novel single-channel IR
precipitation estimation algorithm to provide background precipitation estimates
for merged satellite precipitation datasets. The retrieval has been designed
specifically to work with only a single IR window channel allowing it to be
applied for the full IMERG record starting in 1998 and potentially extended to
the Gridsat B1 record \citep{knapp_gridsat_2011} of three-hourly IR imagery
starting in 1983. The retrieval uses a CNN applied to multiple timesteps of
geostationary observations in order to leverage both spatial and temporal
structure for improved precipitation estimates.

\subsection{Retrieval Accuracy}

This manuscript presents an extensive validation of GPROF-IR precipitation
estimates against independent reference measurements. The evaluation uses
gauge-corrected ground-based radar estimates over CONUS and Korea, dense gauge
network measurements over Austria, and ship-borne disdrometer observations over
the global oceans. Over CONUS and Austria, GPROF-IR yields higher accuracy than
IMERG during GMI overpasses. Over Korea the accuracy of GPROF-IR remains lower
than that of IMERG during GMI overpasses but still greatly surpasses that of the
IMERG V07 IR precipitation fields that are based on the same satellite
observations as GPROF-IR. When evaluation is extended to a full year of
retrievals (thus including IMERG estimates outside of GMI overpasses), GPROF-IR
shows consistently better agreement with ground-based radar estimates for both
hourly and daily accumulations than IMERG V07. Higher accuracy relative to IMERG V07
is also found when comparing precipitation estimates against ship-borne
disdrometer observations. This is largely driven by an improved representation
of precipitation in the tropics, whereas the accuracy in the outside the tropics
remains marginally lower than IMERG V07.

Comparison with IR-based precipitation products from the PERSIANN family
\citep{hong_ccs_2004, nguyen2020persiann, nguyen_punet_2026} further shows that
GPROF-IR improves upon existing IR-based retrieval algorithms despite relying on
the same geostationary satellite observations. GPROF-IR also provides
substantially and consistently more accurate precipitation estimates than ERA5
except for daily accumulations of winter-time precipitation. However, ERA5 may
benefit here from the regional assimilation of precipitation rates from
ground-based radars.

The ability of GPROF-IR to achieve higher accuracy than corresponding IMERG V07
precipitation estimates during GMI overpasses over CONUS and Austria is notable
because GMI is the reference radiometer of the GPM constellation and provides
some of the highest-quality PMW precipitation estimates available to IMERG.
Given that geostationary IR observations are available every 30 min over much of
the globe, the ability of GPROF-IR to come closer to and under certain conditions
surpass the accuracy of much less frequently available PMW retrievals represents
an important advance for satellite precipitation monitoring. It should be noted,
however, that the next operational version of GPROF, GPROF V08, will employ a
related CNN-based retrieval approach that has also been shown to substantially
improve PMW precipitation retrievals.

GPROF-IR also achieves higher effective resolution than IMERG V07 during GMI
overpasses over CONUS and an effective resolution of 1.0$^\circ$ over Korea far
surpassing the effective resolution of the IMERG V07 IR precipitation estimates.
This result may seem at odds with smooth precipitation features retrieved in the
case study of Typhoon Khanun in Fig.~\ref{fig:case_study_1}. The IMERG V07 IR
retrievals exhibit more finescale structure than the GPROF-IR retrievals.
However, the analysis of the spectral coherence suggests that this finescale
variability does not correlate with the finescale features in the ground-based
reference data and is therefore unlikely to represent actual precipitation
features. In fact, smoothing the IMERG IR precipitation fields for this case
study decreases the MSE and increases the correlation coefficient indicating
that removing the finescale structure from the retrievals improves agreement
with the ground-based reference data (Fig.~\ref{fig:case_study_1_smoothed}).
Further increasing the effective resolution of the retrievals will require
increasing the information content of the input observations, which could be
achieved ingesting additional satellite observations. Alternatively, finescale
structure may be recovered by applying tailored loss functions
\citep{guilloteau2023} or probabilistically using generative approaches
\citep{li2026video}.

\subsection{Retrieval Design}

The design of GPROF-IR aims to maximize (1) the amount and quality of the
training data and (2) the information content of the retrieval input data. A
prerequisite for allowing GPROF-IR to yield higher-accuracy estimates than
IMERG is that GPROF-IR uses reference data from the GPM combined DPR and GMI
retrievals and GPROF V08 GMI, both of which yield more accurate precipitation
estimates than IMERG V07. Due to their empirical nature, neural-network-based
retrievals are unlikely to yield more accurate precipitation estimates than the
data they are trained on (although they may potentially reduce random errors
in the reference data). Therefore, retrievals trained on IMERG (or ERA5)
precipitation will remain limited by the accuracy of the reference data.

Furthermore, GPROF-IR uses a single, large training dataset comprising all
available reference precipitation estimates from 2015 until 2018 instead of
training different models for different contexts. Compared to the training data
used for the GPROF-NN PMW retrieval, which uses only reference estimates from
the central part of the DPR swath, the GPROF-IR training data is about eight
times larger for the 2BCMB-based configuration and 16 times larger for the
GPROF-V08-GMI-based configuration.

The GPROF-IR configurations trained using GPROF V08 GMI data consistently
outperformed the configuration trained on 2BCMB data. The training data used to
train the GPROF V08 GMI retrieval implements several ad hoc corrections designed
to address known limitations of the 2BCMB data but nonetheless remains largely
based the 2BCMB data itself. It therefore seems unlikely that improved
performance of the GMI-based GPROF-IR configurations is due to higher accuracy
of the GPROF V08 GMI reference data. Instead, the larger volume of available
GPROF V08 GMI training data may contribute to the observed improvement. This
suggests that future retrieval development could benefit from further expanding
the size of the training dataset.

To maximize the information content in the retrieval input data, GPROF-IR uses
the gridded IR observations at their native resolution aiming to conserve
small-scale cloud texture information in the observations. A positive impact of
maximizing the resolution of the input observations has already been found in
\citet{pfreundschuh_improved_2022}.

\subsubsection{Temporal Information}

In addition to spatial information contained in the geostationary imagery,
GPROF-IR explored the use of temporal information to help constrain the
precipitation estimates. The incorporation of temporal information into
satellite precipitation retrievals has a long history starting with the earliest
methods conceived for satellite based precipitation estimation
\citep{griffith1976rainfall}. However, most current operational geostationary
retrieval algorithms do not make use of any temporal information
\citep{kuligowski_02, nguyen2020persiann}.

In principle, the advent of deep learning should have made the exploitation of
temporal information in geostationary observations easier. However, among the
published deep-learning-based algorithms that exploit temporal information in
geostationary observations not all clearly isolate the effect of temporal
information \citep{cannon2026deep}, are limited to single pixels
\citep{upadhyaya2022exploring}, or even report accuracy degradation when
incorporating temporal information \citep{berthomier2023espresso}. Several
recent algorithms that demonstrate improvements from the incorporation of
temporal information estimate hourly precipitation accumulations instead of
quasi-instantaneous estimates for which the temporal evolution is more directly
relevant \citep{johncox2025, li2026satellite, liu2026exploring}. Previous
studies thus don't provide a consistent picture whether and to what extent
temporal information in geostationary imagery can constrain instantaneous
precipitation estimates.

There are also substantial differences between the techniques used to implement
retrievals using temporal information. These range from single-pixel retrievals
\citep{upadhyaya2022exploring, liu2023}, spatially-resolved retrievals with
explicit handling of the time dimension \citep{cannon2026deep, ouallouche2026}
to simple stacking along the input-feature dimension \citep{berthomier2023espresso}.
There is thus neither a consensus on how to best implement retrieval techniques
leveraging temporal information.

GPROF-IR follows the approach of stacking the input images along the temporal
dimension and does not introduce specific modules for merging the temporal
information. The difference to the comparable approach approach pursued by
\citet{berthomier2023espresso} is the sequential training of the models
ingesting temporal information. The sequential training was adopted because
multi-step models trained from scratch would yield lower retrieval accuracy than
single-step models under identical training conditions. This behavior is similar
to the reported degradation reported by \citet{berthomier2023espresso} when
observations from multiple time steps are used. This suggests that sequential
training may be an effective approach to address these difficulties.

Our results show moderate improvements from the integration of temporal
information into the retrieval. However, these improvements are robust and
generalize to validation against independent precipitation estimates
over both land and ocean. Our analysis  largely isolates the impact of the
integration of temporal information with the limitation that the temporal models
have undergone longer training due to the initialization from the previous
training stage. While the accuracy gains are moderate, temporal information is
readily available from geostationary sensors and its ingestion thus presents a
straightforward way to improve precipitation estimates from geostationary
sensors.

An analysis of the spatial distribution of the improvements achieved by the
incorporation of temporal information are shown in
Fig.~\ref{fig:test_results_improvements}. The improvements in MAE and
correlation coefficient are robustly positive with only the MSE showing limited
regions of accuracy degradation. The results are noisy but regions with
increased improvement are discernible over northern Africa, central Asia, the
western US, as well as the sub-tropical subsistence zones off the coasts of
South America, Africa, and Australia. Improvements over the western US are also
found for hourly accumulations over CONUS (Fig.~\ref{fig:conus_temporal_impact})
Given that these regions are characterized by a relatively larger contribution
of shallow precipitation, we speculate that temporal evolution benefits
precipitation regimes with comparably weak radiative signatures and where the
brightness temperature is less directly related to surface precipitation.

Although moderate, the improvements afforded by the ingestion of temporal
information has proven robust in our analysis. An important feature of GPROF-IR
that the multi-timestep retrievals are trained using quasi-instantaneous
training data from global satellite-based reference estimates. This is an
important difference to other retrievals that often rely on reference estimates
from ground-based radar \citep{upadhyaya2022exploring, johncox2025,
cannon2026deep, li2026satellite, ouallouche2026} or station data
\citep{liu2023}. Since ground-based radar and station data are generally
regionally limited it remains unclear how well these estimates generalize
outside their training domains.

\subsubsection{Integration into IMERG V08}

\begin{figure}[htbp]
 \centerline{\includegraphics[width=37pc]{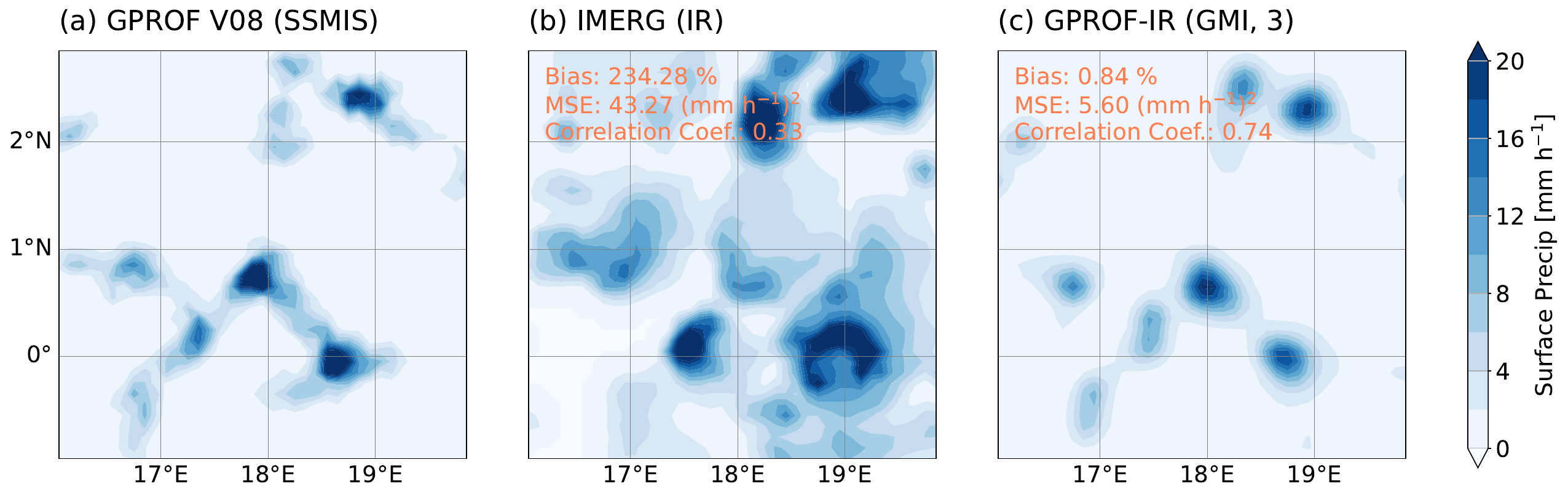}}
 \caption{
Detailed view of the convective precipitation cells during the SSMIS overpass
occurring at 15:00 UTC shown in Fig. 2. Panel (a) shows the estimates from GPROF
V08, which will be used in IMERG V08; panel (b) shows the IMERG V07 IR
precipitation estimate; panel (c ) shows the corresponding estimates of the
GPROF-IR (GMI, 3) retrieval that will be used to produce precipitation estimates
for IMERG V08
 }\label{fig:evolution_2}
\end{figure}

Because of its capability to greatly improve the accuracy of IR-only
precipitation estimates, the GPROF-IR (GMI, 3) retrieval will be used to produce IR
precipitation estimates for next version of IMERG -- IMERG V08.
Figure~\ref{fig:evolution_2} revisits the convective precipitation system from
Fig.~\ref{fig:evolution} displaying the GPROF V08 PMW precipitation estimates
that will be used in IMERG V08, the current IR precipitation estimates from
IMERG V07, and the new GPROF-IR (GMI, 3) retrieval results. In contrast to the
IMERG V07 IR precipitation estimates, GPROF-IR produces a precipitation field
that exhibits consistent precipitation structures compared to the PMW estimates.
This suggests that the GPROF-IR retrieval may help to improve the
overall accuracy of IMERG V08 but also improve the representation of the
temporal evolution of precipitation systems.

Nonetheless, the final impact of GPROF-IR on IMERG V08 remains an open question. As
shown in Fig.~\ref{fig:availability}, IR precipitation only contributes weakly to
the merged precipitation estimates in IMERG V07. This is because the error-aware
merging scheme assign low weights to the comparably inaccurate conventional IR
precipitation fields. Given the greatly improved accuracy of GPROF-IR, it is likely
that the contribution of the IR-only estimates will increase in IMERG V08. However,
quantifying the final impact will only be possible after integration into IMERG V08,
which will involve additional evaluation of the GPROF-IR retrievals against the PMW
retrievals of the GPM constellation.

\section{Conclusions}

The GPROF-IR precipitation retrieval algorithm is a novel single-channel IR
retrieval that provides substantially improved precipitation estimates compared
to currently available single-channel IR approaches. Validation against independent
ground-based reference measurements over Austria and CONUS demonstrate that
GPROF-IR outperforms  widely used precipitation products across
timescales ranging from quasi-instantaneous estimates to daily accumulations.
GPROF-IR was developed to provide background precipitation estimates for the
upcoming IMERG V08 merged precipitation product. However, because the retrieval
relies exclusively on a single IR channel, its applicability extends beyond this
operational role and may enable the generation of consistent precipitation
estimates from the geostationary IR record dating back to 1983.

In addition to its planned operational use in IMERG V08, the GPROF-IR retrievals
and their implementation are openly available to the scientific community
\citep{pfreundschuh_gprofir_2026}. Because the retrieval requires only IR
observations as input, it can be readily applied to the CPC gridded IR record
without additional ancillary data. Its quasi-global and long-term applicability,
as well as demonstrated accuracy also make GPROF-IR a useful baseline for the
development and evaluation of future satellite precipitation retrievals.

\clearpage
\section*{Acknowledgments}

The work of SP and CDK was supported by NASA grant 80NSSC22K0604.

This study uses modified Copernicus Climate Change Service information. Neither the European Commission nor ECMWF is responsible for any use that may be made of the Copernicus information or data it contains.

We acknowledge the feedback from three anonymous reviewers that helped us to improve the manuscript.

\section*{Data Availability Statement}

The NOAA CPC 4-km gridded IR observations are available from \citet{ncep_cpc_mergir_v1}. The GPM 2BCMB data are available  from \citet{cmb_data}. The SatRain dataset is available from \citet{satrain_dataset}. The OceanRAIN data are available from \citet{ocean_rain_data}. The ERA5 data are available from \citet{c3s_era5_single_levels}. The GPCP data are available from \citet{gpcp_v33_precip}.


\clearpage

\appendix

\section{Supplementary Test Results}

\subsection{Spatial Distribution of Accuracy Metrics}

\begin{figure}[htbp]
 \centerline{\includegraphics[width=37pc]{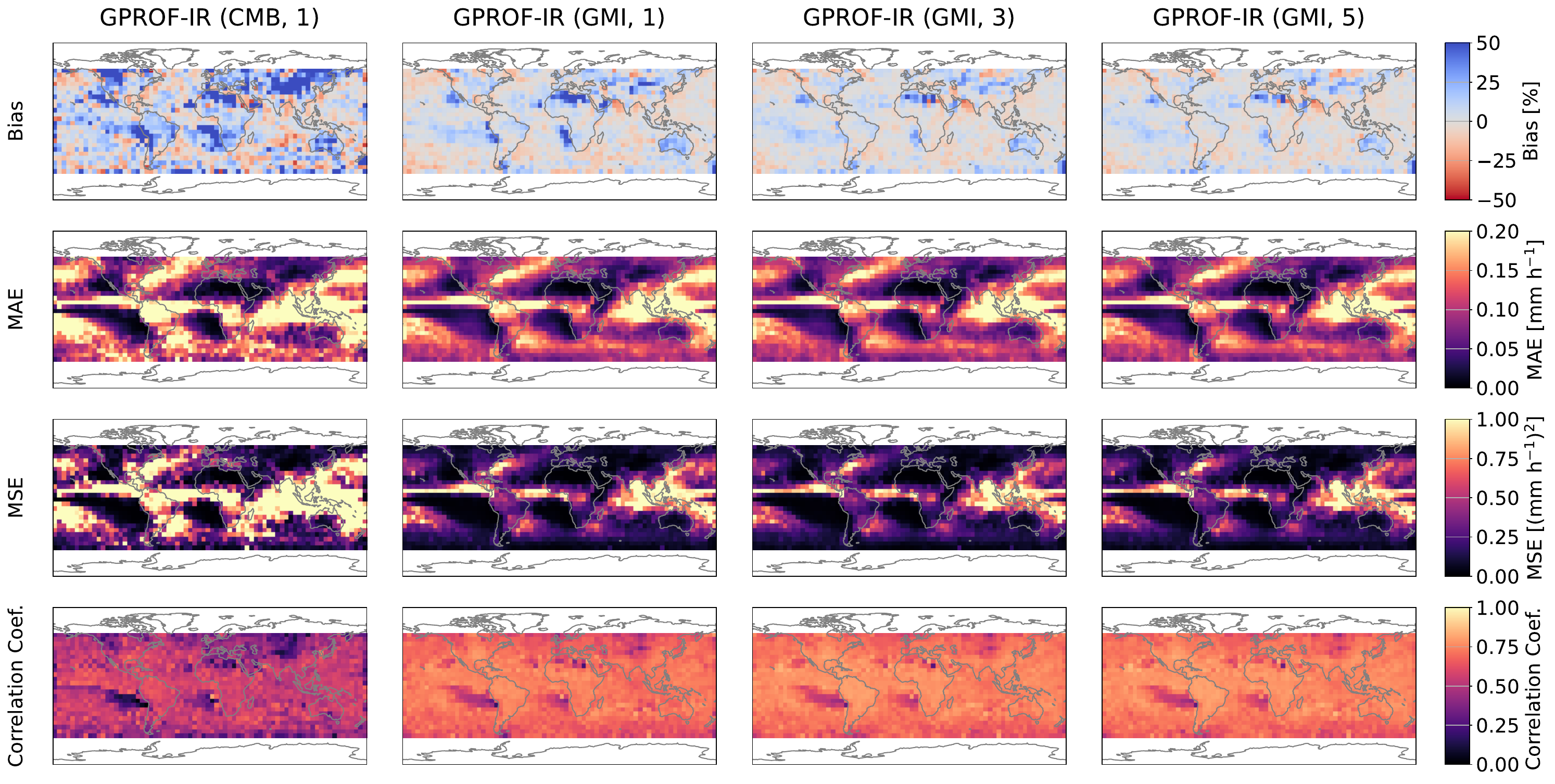}}
 \caption{Spatial distributions of bias (first row), mean absolute error (second row),
   mean squared error (third row), and correlation coefficient (fourth row). Columns 1 -- 4
   show the results for the GPROF-IR (CMB, 1), GPROF-IR (GMI, 1), GPROF-IR (GMI, 3), and GPROF-IR (GMI, 5)
   configurations, respectively.
 }\label{fig:test_results_spatial}
\end{figure}

Figure~\ref{fig:test_results_spatial} shows the spatial distributions of the bias, MAE, MSE, and correlation
coefficient calculate using the temporally independent testing data from 2022.

\clearpage

\subsection{Impact of Temporal Information}

\begin{figure}[htbp]
 \centerline{\includegraphics[width=37pc]{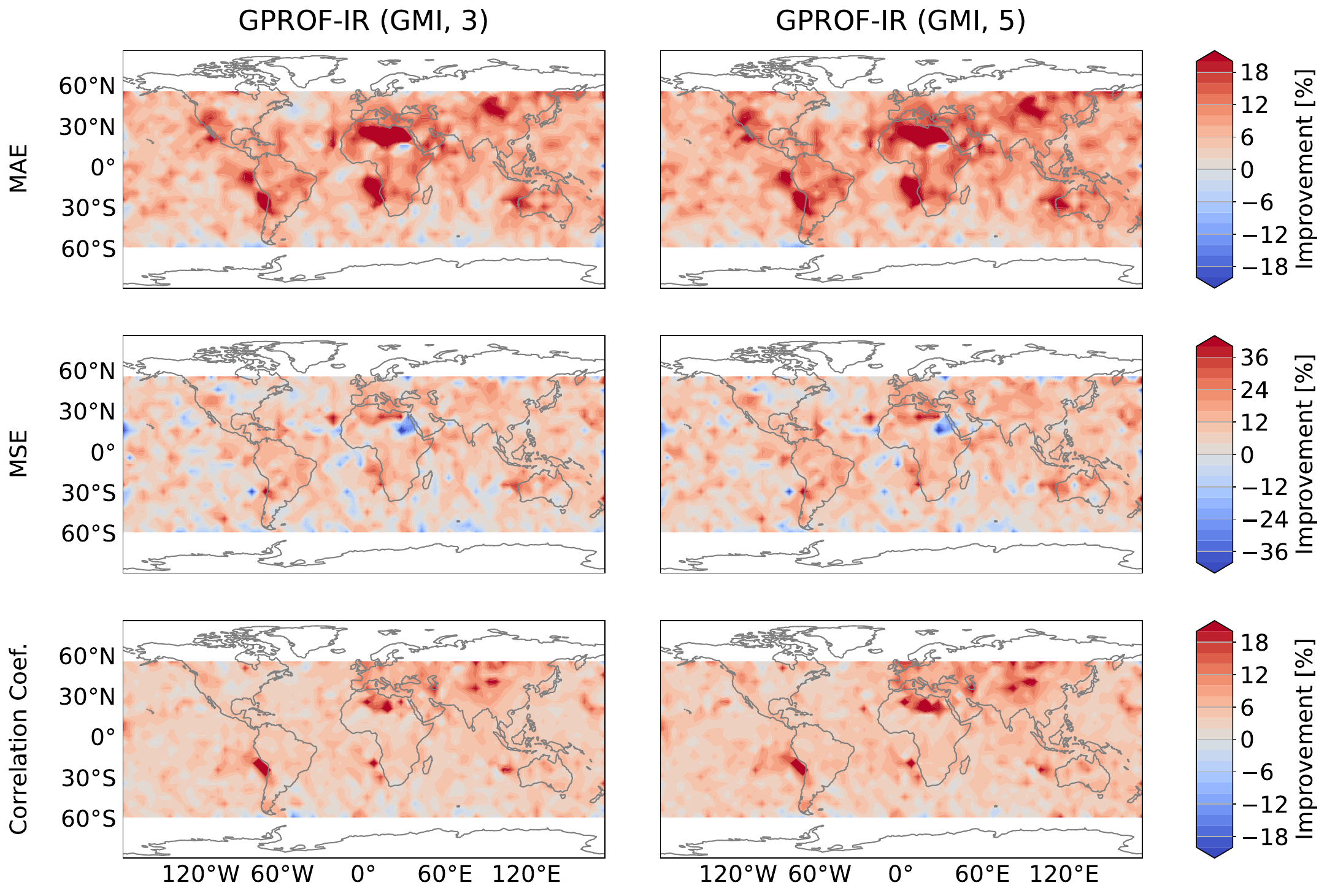}}
 \caption{Spatial distribution of the accuracy improvement for the three-timestep retrieval GPROF-IR (GMI, 3)  and five-timestep retrieval GPROF-IR (GMI, 5) relative to the single-step retrieval GPROF-IR (GMI, 1). The first row shows improvements in the mean absolute error (MAE), the second row for the mean squared error (MSE), and the third row for the correlation coefficient.}
 \label{fig:test_results_improvements}
\end{figure}

Figure~\ref{fig:test_results_improvements} shows the spatial distribution of the relative improvements in MAE, MSE, and correlation for the multi-timestep GPROF-IR (GMI, 3) and GPROF-IR (GMI, 5) retrievals compared to the single-timestep GPROF-IR (GMI, 1) retrieval.

\clearpage

\section{Impact of Spatial Smoothing on IMERG IR Precipitation}

\begin{figure}[htbp]
 \centerline{\includegraphics[width=37pc]{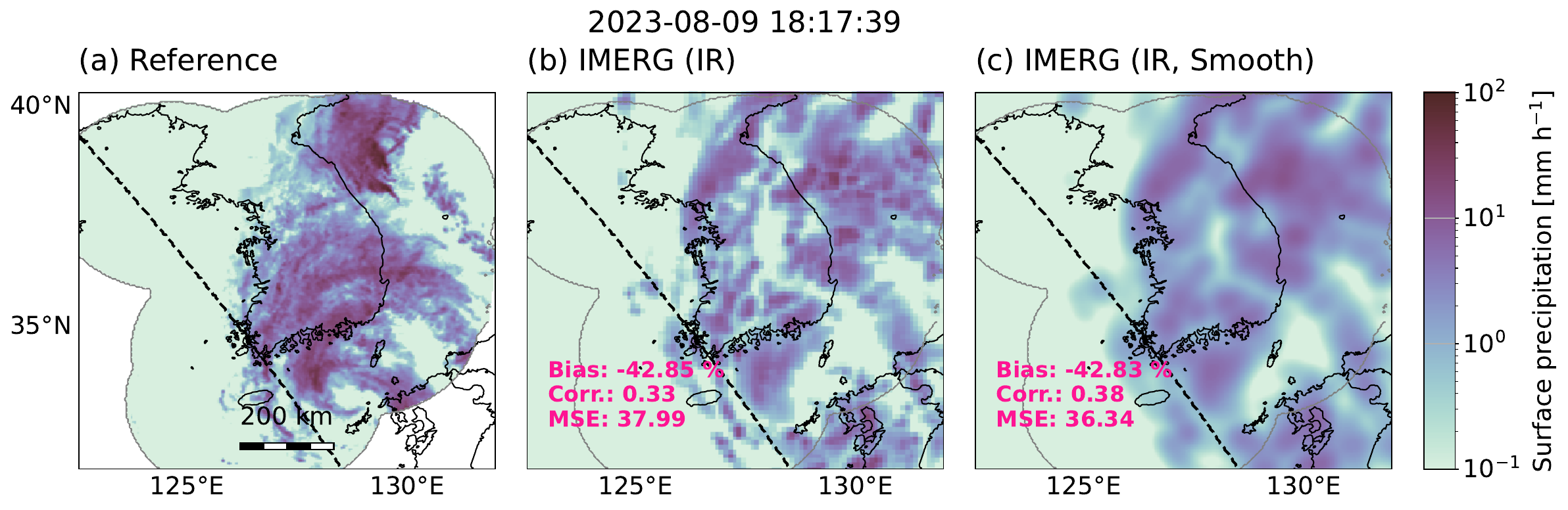}}
 \caption{Smoothed IMERG IR precipitation fields compared to the original IMERG
IR fields and the ground-based reference for the case study of Typhoon Khanun
shown in Fig.~\ref{fig:case_study_1}}\label{fig:case_study_1_smoothed}.
\end{figure}

Figure~\ref{fig:case_study_1_smoothed} shows the impact of removing small-scale
variability from the IMERG IR precipitation fields. The figure shows the
original IMERG V07 IR precipitation field and the same field smoothed using a Gaussian
kernel with a full-width-at-half-maximum of 0.2$^\circ$. The Gaussian smoothing
acts as a low-pass filter removing small-scale variability. The increase in
correlation and decrease in MSE indicates that the removed small-scale
variability does not correlate with small-scale features in the radar-based
reference fields. This is consistent with the spectral coherence curves shown in
Fig.~\ref{fig:spectral_coherence} that lower coherence with the reference data
than the GPROF-IR retrievals despite exhibiting more small-scale variability.

\clearpage

\section{Validation of Hourly Accumulations over CONUS}

\subsection{Spatial Distribution of Accuracy Metrics}

\begin{figure}[htbp]
 \centerline{\includegraphics[width=37pc]{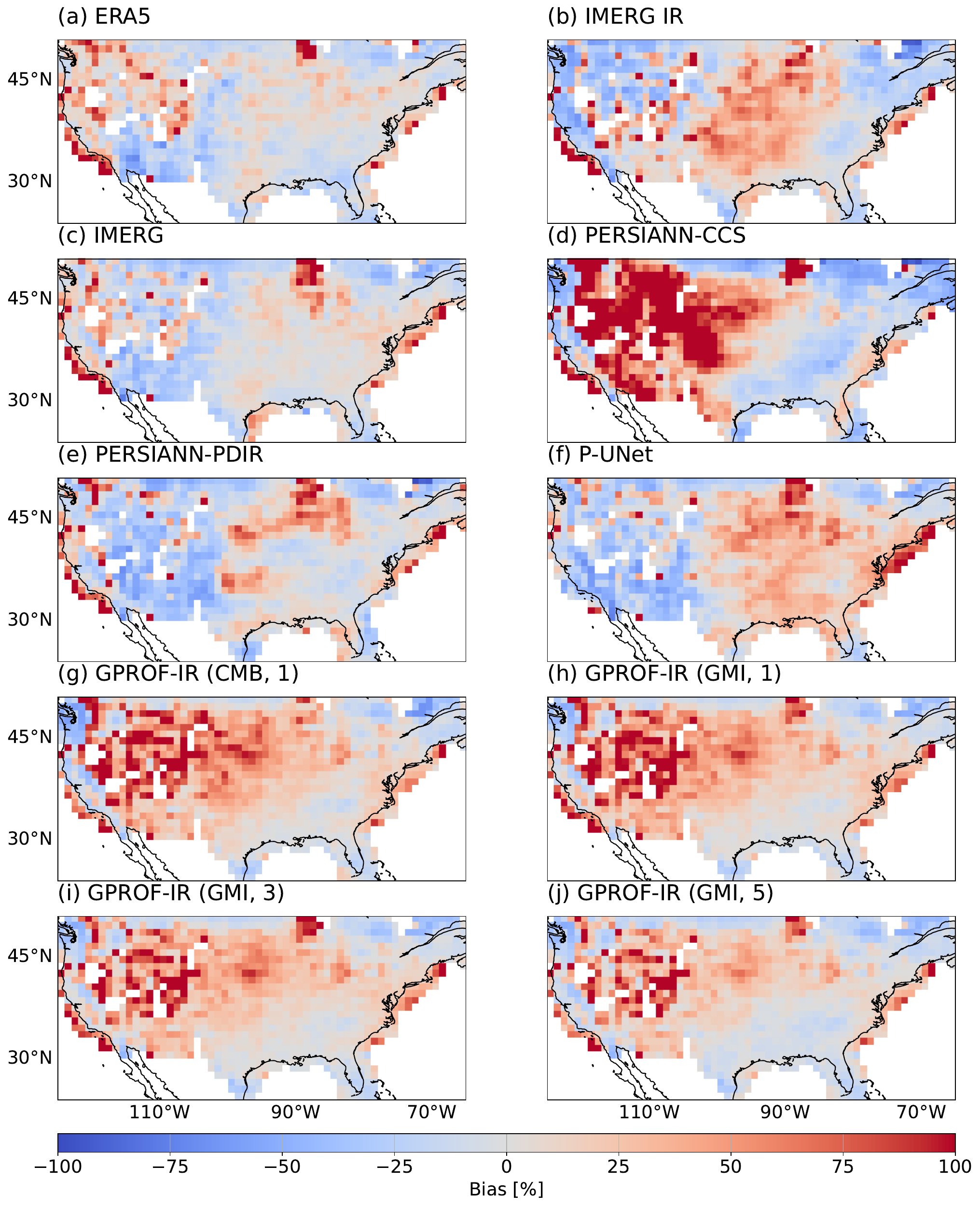}}
 \caption{Spatial distribution of retrieval biases for hourly accumulated precipitation during 2022 compared to gauge-corrected ground-based radar estimates from MRMS.}\label{fig:conus_spatial_bias}
\end{figure}

\begin{figure}[htbp]
 \centerline{\includegraphics[width=37pc]{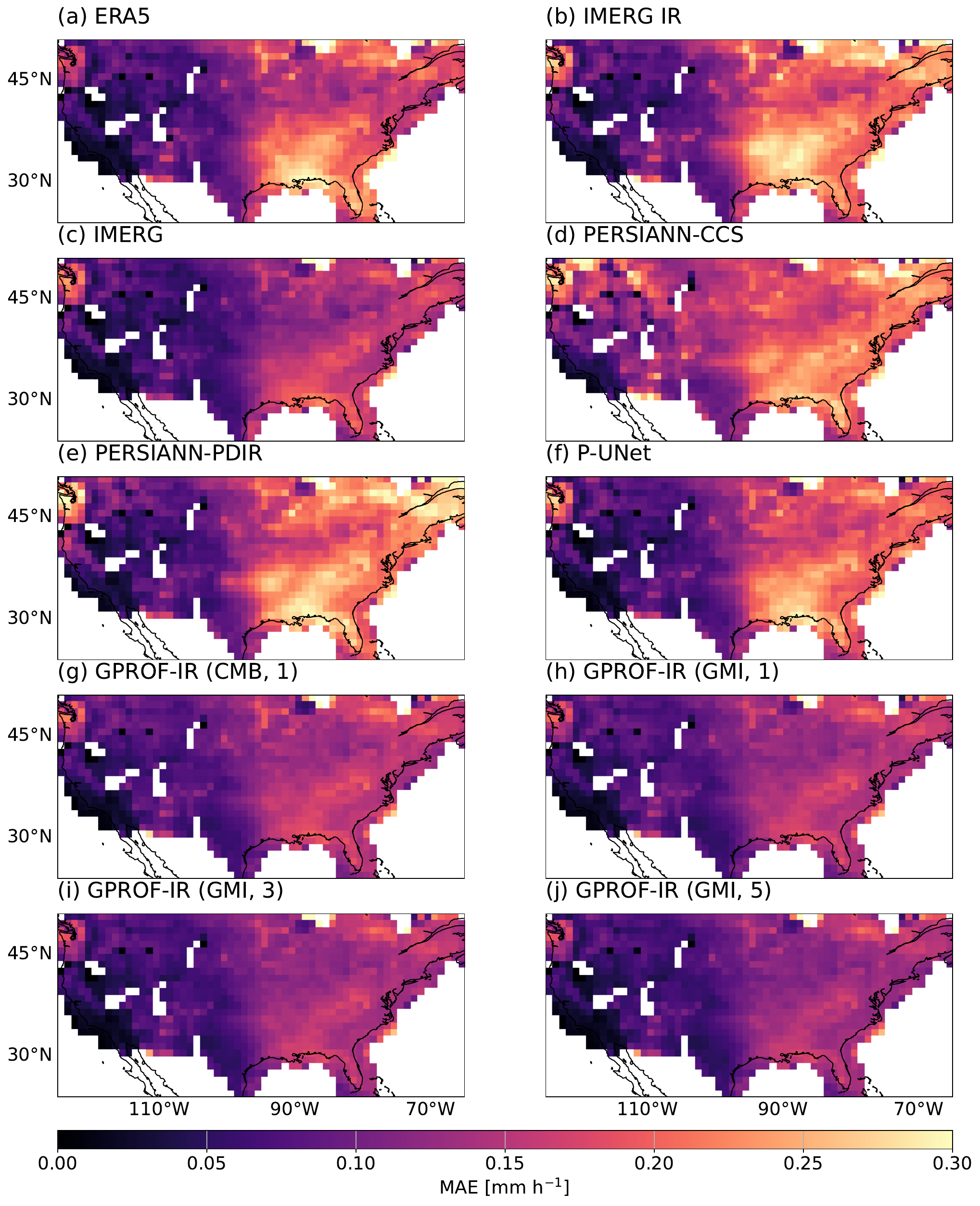}}
 \caption{Spatial distribution of the mean absolute error for hourly accumulated precipitation during 2022 compared to gauge-corrected ground-based radar estimates from MRMS.}\label{fig:conus_spatial_mae}
\end{figure}

\begin{figure}[htbp]
 \centerline{\includegraphics[width=37pc]{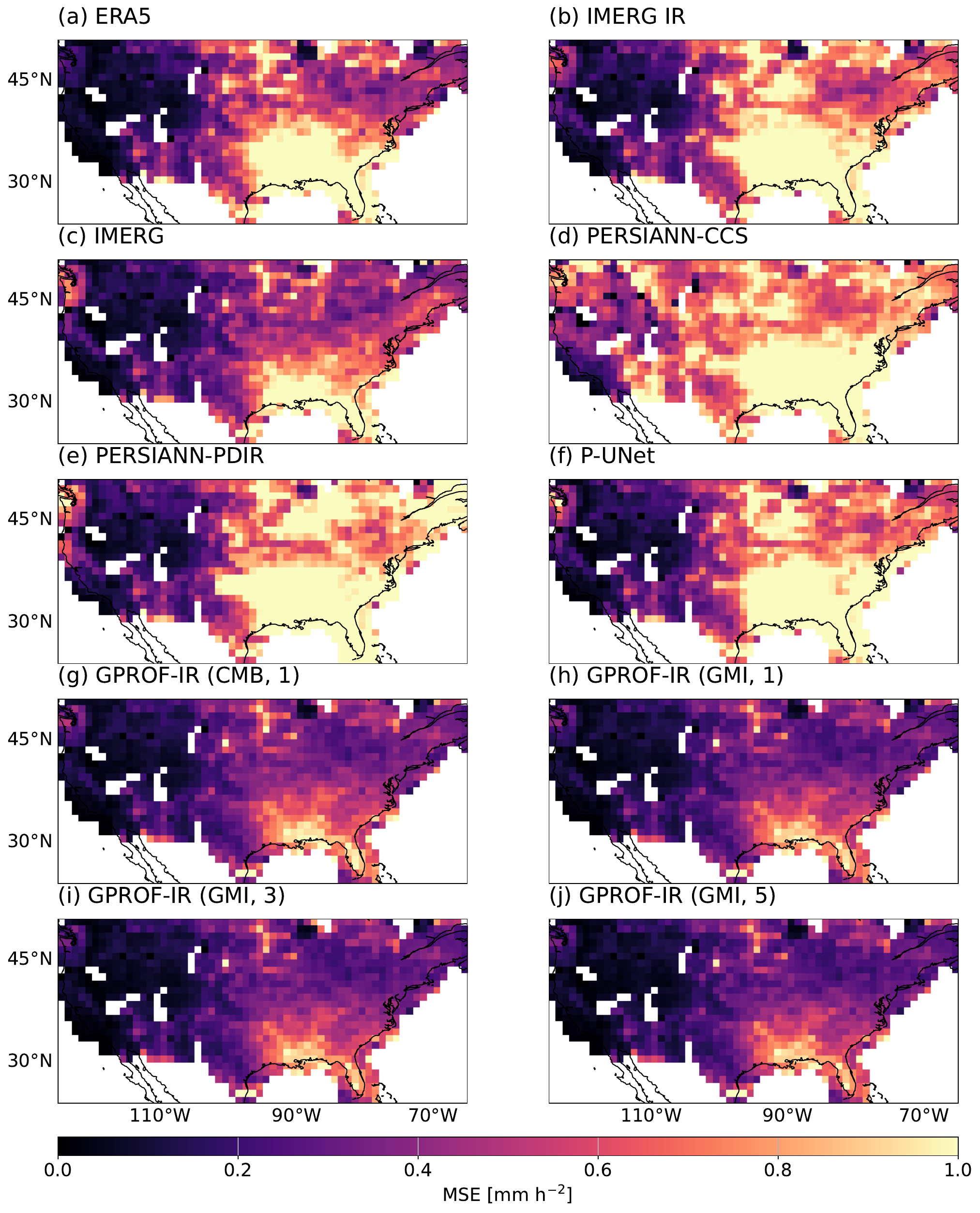}}
 \caption{Spatial distribution of the mean squared error for hourly accumulated precipitation during 2022 compared to gauge-corrected ground-based radar estimates from MRMS.}\label{fig:conus_spatial_mse}
\end{figure}

\begin{figure}[htbp]
 \centerline{\includegraphics[width=37pc]{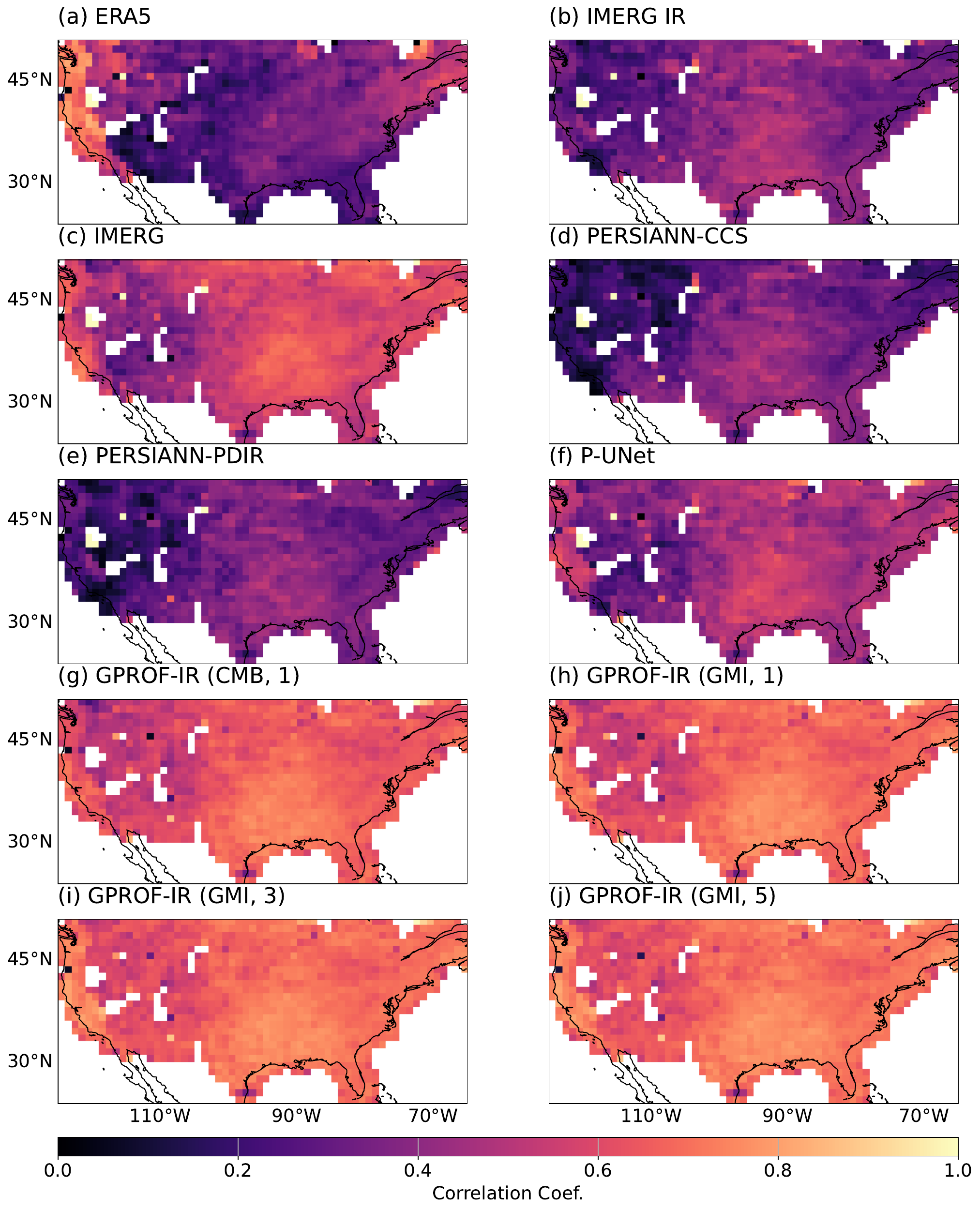}}
 \caption{Spatial distribution of the correlation coefficient for hourly accumulated precipitation during 2022 compared to gauge-corrected ground-based radar estimates from MRMS.}\label{fig:conus_spatial_corr}
\end{figure}

Figures~\ref{fig:conus_spatial_bias} -- \ref{fig:conus_spatial_corr} display the
spatial distributions of the bias, MAE, MSE, and correlation coefficient for the
validation of hourly precipitation accumulations validated using gauge-corrected
ground-based radar estimates from MRMS.

\clearpage

\subsection{Impact of Temporal Information}

\begin{figure}[htbp]
 \centerline{\includegraphics[width=37pc]{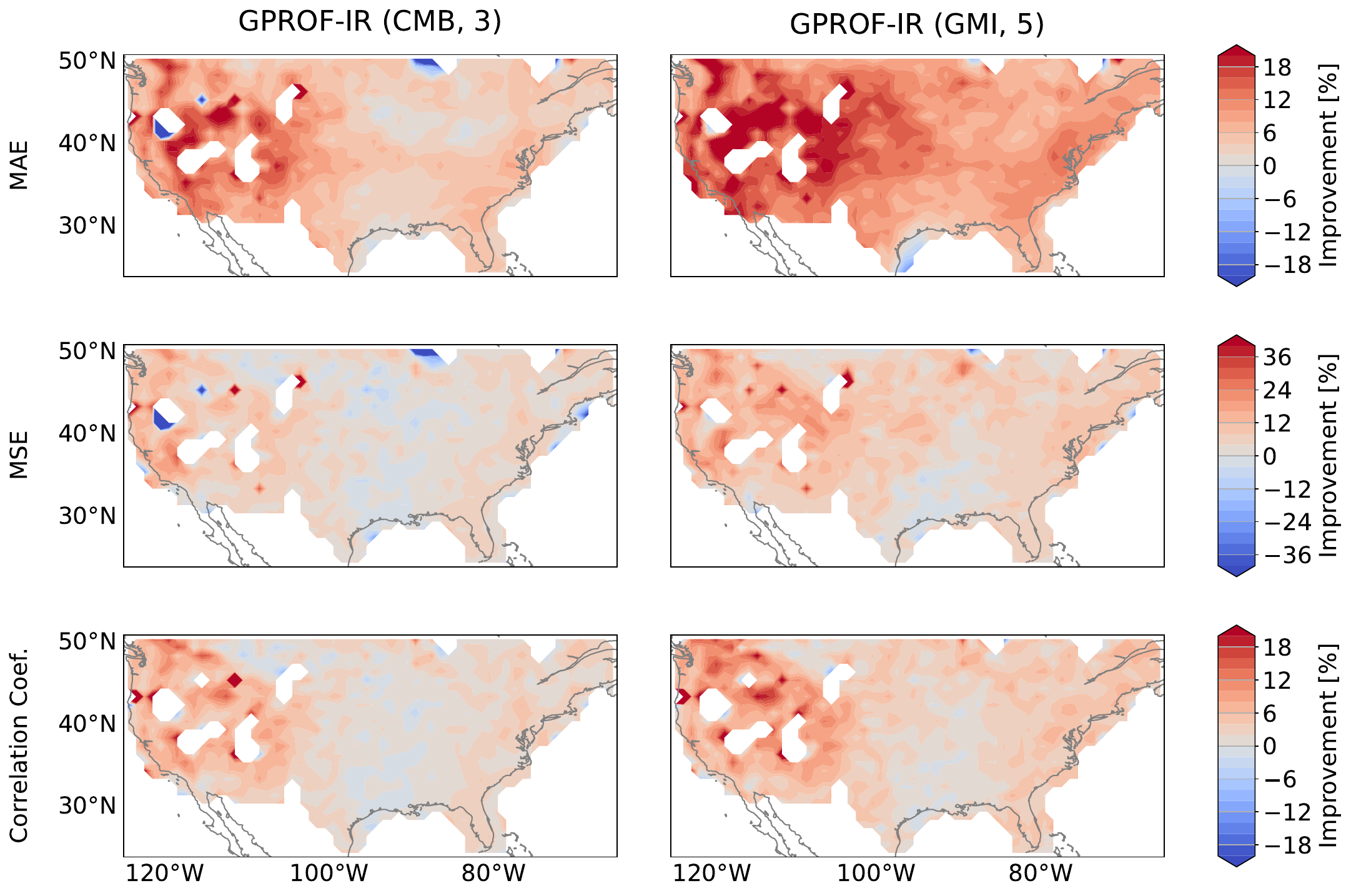}}
 \caption{Relative improvements in the validation accuracy of hourly precipitation
accumulations from GPROF-IR retrievals using three (GPROF-IR (GMI, 3)) and five
(GPROF-IR (GMI, 5)) consecutive half-hourly, single-channel geostationary IR
observations compared to the single-timestep retrieval (GPROF-IR (GMI, 1)).
Validation is performed against independent gauge-corrected ground-based radar
precipitation estimates.}\label{fig:conus_temporal_impact}
\end{figure}

\clearpage


\bibliographystyle{plainnat}
\bibliography{references}

\end{document}